\documentclass[twocolumn]{aastex631}

\usepackage{threeparttable} 
\usepackage{graphicx}       
\usepackage{xcolor}         
\usepackage{subfigure}
\usepackage{booktabs}
\usepackage{mhchem}
\usepackage{hyperref}
\usepackage{wrapfig}
\usepackage{sidecap}
\usepackage{float}
\usepackage{placeins}
\usepackage{capt-of}

\newlength{\subfigwidth}
\newcommand{\scalefactor}{0.56}

\begin{document}

\begin{center}
    \Large\textbf{The Clear Sky Corridor: \\ Insights Towards Aerosol Formation in Exoplanets\\ Using an AI-based Survey of Exoplanet Atmospheres}
\end{center}

\begin{center}
    Reza Ashtari$^{1}$, Kevin B. Stevenson$^{1}$, David Sing$^{2}$, Mercedes L\'opez-Morales$^{3}$, Munazza K. Alam$^{3}$, \\
    Nikolay K. Nikolov$^{3}$, Thomas M. Evans-Soma$^{4,5}$\\
    
    \vspace{0.2cm}
    
    \textit{
        $^{1}$Johns Hopkins University Applied Physics Laboratory, 11100 Johns Hopkins Rd., Laurel, MD 20723, USA\\
        $^{2}$Johns Hopkins University, 3400 N. Charles St., Baltimore, MD 21218, USA\\
        $^{3}$Space Telescope Science Institute, 3700 San Martin Drive, Baltimore, MD 21218, USA\\
        $^{4}$School of Information and Physical Sciences, University of Newcastle, Callaghan, NSW, Australia\\
        $^{5}$Max-Planck-Institut für Astronomie, Heidelberg, Germany
    }
\end{center}

\vspace{0.05cm}
\begin{center}
    \textit{Accepted to AJ}
\end{center}

\vspace{-0.5cm}

\begin{abstract}
Producing optimized and accurate transmission spectra of exoplanets from telescope data has traditionally been a manual and labor intensive procedure. Here we present the results of the first attempt to improve and standardize this procedure by using artificial-intelligence based (AI-based) processing of light curves and spectroscopic data from transiting exoplanets observed with the Hubble Space Telescope's (HST) Wide Field Camera (WFC3) instrument. We implement an AI-based parameter optimizer that autonomously operates the Eureka! pipeline to produce homogeneous transmission spectra of publicly available HST WFC3 datasets, spanning exoplanet types from hot Jupiters to sub-Neptunes. Surveying 42 exoplanets with temperatures between 280 -- 2580 K, we confirm modeled relationships between the amplitude of the water band at 1.4 $\micron$ of hot Jupiters and their equilibrium temperatures. We also identify a similar, novel trend in Neptune/sub-Neptune atmospheres, but shifted to cooler temperatures. Excitingly, a planet mass versus equilibrium temperature diagram reveals a “Clear Sky Corridor", where planets between 700 -- 1700 K (depending on the mass) show stronger 1.4 {\micron} \ce{H2O} band measurements. This novel trend points to metallicity as a potentially-important driver of aerosol formation. With HST sculpting this foundational understanding for aerosol formation in various exoplanet types ranging from Jupiters to sub-Neptunes, we present a compelling platform for the James Webb Space Telescope (JWST) to discover similar atmospheric trends for more planets across a broader wavelength range.
\end{abstract}


\section{Introduction} 
\label{sec:intro}

Twenty years after its first transit observation, Hubble's impact on exoplanet research is monumental. Hubble provided the first detailed looks at transiting exoplanets \citep{Brown2001} and exoplanet atmospheres \citep{Charbonneau2002}, pioneering exoplanet characterization. Two decades later, HST remains a leader in measuring exoplanet spectra, uniquely capable of characterizing atmospheres from those of super-Earths to gas giants at UV to near-IR wavelengths. HST has surveyed the atmospheres of dozens of gas giants \citep{Sing2016} and detected spectral features in smaller planets \citep{Wakeford2017, Benneke2019}. Discoveries include \ce{H2O}, \ce{Na}, and \ce{K} in numerous planets \citep[e.g.][]{Deming2013, Fu2017}, high-temperature clouds and hazes \citep{Pont2008}, atmospheric escape \citep{Vidal2003}, and diverse water abundance / metallicity measurements \citep{Kreidberg2014a, Wakeford2017, Spake2021}.

Most HST exoplanet spectra come from transmission spectroscopy, which requires high spectrophotometric precision (SNR $\sim$ 10,000). Despite not being designed for time-series observations, HST instruments have achieved precisions down to 20 ppm \citep{Charbonneau2002, Kreidberg2014a, Wakeford2017}. HST has dedicated over 1000 orbits to exoplanet atmospheres, with more than 68 exoplanets having near-IR data and 35 with both optical and near-IR data \citep{Nikolov2022}. This rich dataset is full of potential for new discoveries, though it has not yet been uniformly analyzed.

Previous attempts to process this wealth of data and produce homogenous transmission spectra have not always yielded consistent results, largely due to a lack of data optimization protocol when analyzing large surveys \citep{Tsiaras2017, Edwards2023}.
To combat the inconsistencies and difficulties of optimally reducing these huge datasets, a more autonomous method for data optimization must be used to produce reliable homogeneously analyzed spectra. To simplify and standardize the processing of spectroscopic transit observations, we present an autonomous data optimizer for HST WFC3 exoplanet datasets, capable of repeatedly and robustly producing high-quality transit light curves and transmission spectra.

AI-based tools have been used to find transiting exoplanets since at least 2017 \citep{Pearson2017, Schanche2019, Terry2023}, most recently and successfully from Kepler and TESS observations using \texttt{ExoMINER} \citep{Valizadegan2022, Valizadegan2023}. To our knowledge, there have been no attempts to use AI for transmission spectroscopy. To automate the transit reduction and atmospheric characterization processes necessary for the study of exoplanet atmospheres, we offer a novel AI-based solution to \texttt{Eureka$!$}, an open-source, community-developed pipeline for processing HST and JWST exoplanet observations \citep{Bell2022}.

An overview of the \texttt{Eureka$!$} pipeline for processing transit observations is given in \autoref{sec:Eureka}. A review of how AI is implemented towards automating optimization of  transit observations with the \texttt{Eureka$!$} HST pipeline is provided in  \autoref{sec:AI_intro}. Results establishing proof-of-concept for AI-based processing of 20 exoplanet transmission spectra are also discussed in this section. The homogeneously produced spectra from this work are presented and discussed in \autoref{sec:water_findings}. In \autoref{sec:comparative}, we present a large-scale comparative study of \ce{H2O} feature amplitudes of 42 exoplanet spectra and discuss what it tells us about aerosol formation in exoplanet atmospheres. The methods and findings from these sections are summarized in \autoref{sec:Summary}.


\section{Processing JWST \& HST Transits \\with \texttt{Eureka$!$}}
\label{sec:Eureka}

\texttt{Eureka$!$} is a specialized data reduction and analysis pipeline tailored for time-series observations of exoplanets, with a specific emphasis on data from JWST \citep{Bell2022}. As an open-source resource, it provides astronomers with a community-developed and well-documented tool for analyzing the atmospheres of distant worlds. Ultimately, \texttt{Eureka$!$} is a tool for converting raw, uncalibrated FITS images into precise exoplanet transmission and/or emission spectra.

The \texttt{Eureka$!$} pipeline features a modular design consisting of six stages, four of which (Stages 3-6) are used for HST WFC3 transit observations. An itemized description for each of these stages is provided below.

\begin{itemize}
\item Stage 1 - Detector Processing - An optional step that calibrates raw data (converts ramps to slopes for JWST observations).

\item Stage 2 - Data Reduction - An optional step that further calibrates Stage 1 data (performs flat-fielding, unit conversion, etc. for JWST observations).

\item Stage 3 - Data Reduction - Performs background subtraction and optimal spectral extraction on calibrated image data. For spectroscopic observations, this stage generates a time series of 1D spectra.
  
\item Stage 4 - Lightcurve Generation - Using Stage 3 outputs, generates spectroscopic light curves by binning the time series of 1D spectra along the wavelength axis. Optionally removes drift/jitter along the dispersion direction and/or sigma clips outliers.
  
\item Stage 5 - Lightcurve Fitting - Fits the light curves with noise and astrophysical models using different optimization or sampling algorithms.

\item Stage 6 - Spectra Plotting - Displays the planet spectrum in figure and table form using results from the Stage 5 fits.
  
\end{itemize}

Each of these stages is managed by ``Eureka! Control Files" (ECFs) and ``Eureka! Parameter Files" (EPFs). These files guide the pipeline operations, with EPFs specifically adjusting the transit model fit parameters. Additional details are available on \texttt{Eureka$!$} \texttt{ReadTheDocs} page\footnote[1]{https://eurekadocs.readthedocs.io/en/latest/index.html}. \texttt{Eureka$!$} currently provides template ECFs for the MIRI, NIRCam, and NIRSpec instruments on JWST\footnote[2]{NIRISS instrument capability to be added by summer 2025 \\(per private communication)}, as well as for the WFC3 instrument on HST, offering a solid foundation for the analysis of exoplanet atmospheres using high-precision space telescope measurements. However, \texttt{Eureka$!$} is not designed for ``black box" use \citep{Bell2022}; it requires users to carefully adjust numerous settings for each dataset to ensure optimal results.
A full, manual exploration of the entire parameter space is an intractable problem for humans.  Autonomous optimization is the right solution for finding the best settings that minimize the standard deviation in the light curve residuals.

\section{AI-based Optimization \\of Transmission Spectra}
\label{sec:AI_intro}

Each high-precision transit observation made with HST is like a snowflake, completely unique and unlike any other observation \citep{Stevenson2019}. The pointing stability, temperature of the instrument, brightness of the target, exact location and width of the spectrum on the detector, and several other conditions will never be identical between two measurements. 
In the context of using AI towards processing such observations, this means that data sets cannot be used to train a deep learning / machine learning model. When attempting to automate the processing and optimization of such measurements, intelligent decision-making needs to be made in real-time.

To appropriately reduce data from telescope images, produce light curves, and fit transit models to the light curves, over 20 free parameters (a.k.a. variables) need to be optimized.

In multi-dimensional search spaces, a well-designed optimizer will converge towards one of, if not the best solution possible. The criteria for selecting the best value for each variable is defined by a fitness score. In this application, we use the median absolute deviation (MAD) values of light curves and the $\chi^2$ values of model fits to assess the quality of the light curves produced, and how well the transit models are being fitted to the measurement data.

\begin{figure}[!]
    \centering
    \includegraphics[width=1.0\linewidth, trim={2cm 3cm 0cm 4.6cm},clip]{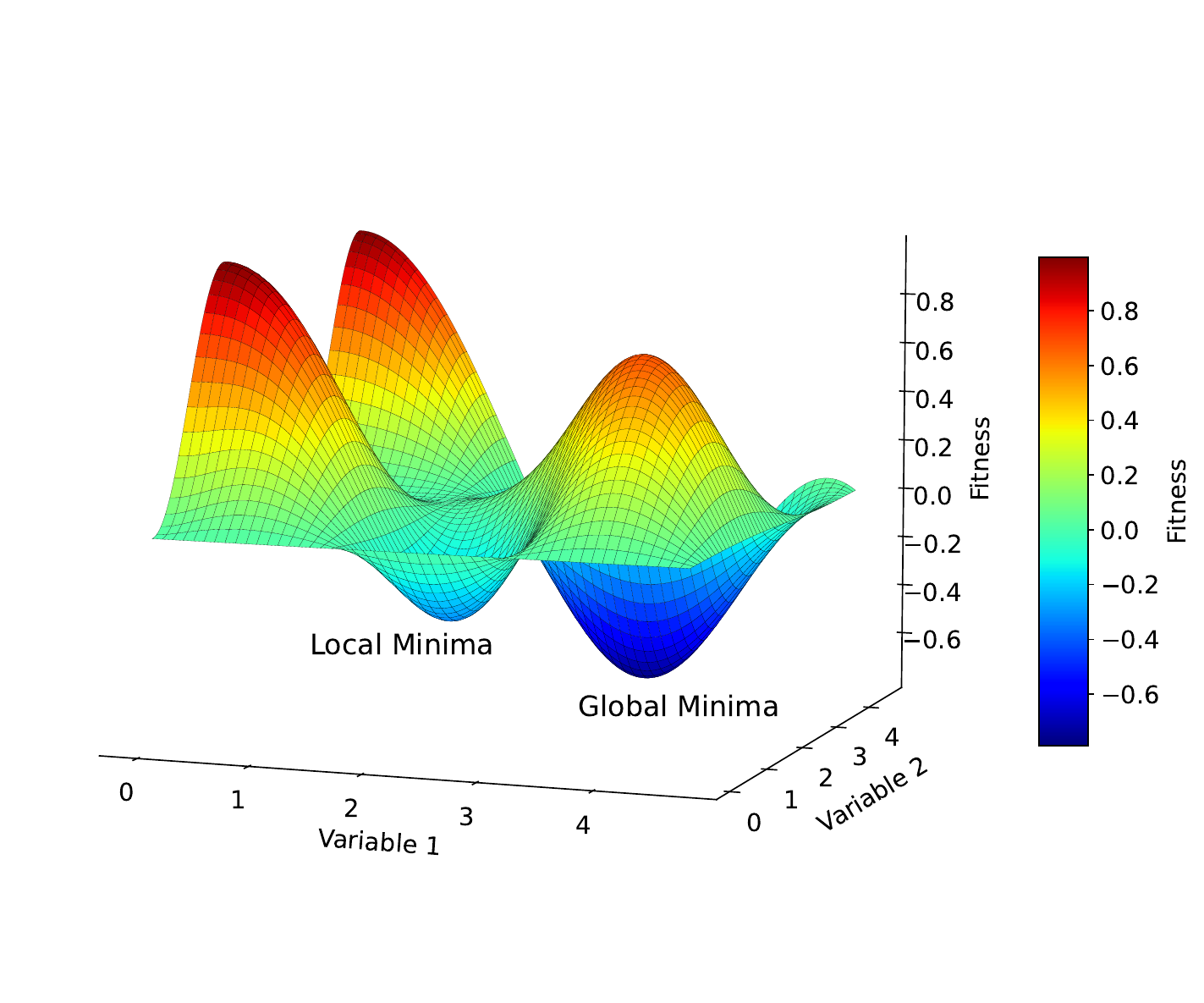} 
    \caption{A visualization of local minimum vs. global minimum. In AI-based optimization, the solution should converge to a minimum by design. The type of AI used for optimization will determine which type of minimum is achievable. Global minima are the best-solutions in the entire search-space. Local minima can be, but are not necessarily the best solutions possible \citep{Charbonneau1995, Rahmat1999}. In this example, a search space for two simultaneously-solved variables is shown in the X and Y axes. The fitness score, shown on the Z-axis, is based on application-specific criteria. Here, lower fitness scores are more optimal solutions.}
    \label{fig:local_global_minima}
\end{figure}

The best fit solutions in a multi-variable search space are located in regions referred to as minima. \autoref{fig:local_global_minima} illustrates a simplified case of two-variable optimization for conceptual purposes. The regions with the best fit solutions are referred to as global minima, while other regions with fit solutions are referred to as local minima. 

The most brute-force, but guaranteed method of converging to a global minimum is to perform a grid search across all values possible for all variables, simultaneously. With over 20 free parameters to solve for, this approach is too computationally-expensive.

Requiring a more pragmatic approach, we consider two less-expensive optimization methods: (1) parametric optimization and (2) genetic optimization \citep{Charbonneau1995, Rahmat1999}.

\begin{figure*}[!]
    \vspace{-1cm}
    \centering
    \begin{minipage}{\textwidth}
        \centering
        \includegraphics[width=0.7\linewidth, trim={2cm 0cm 1.5cm 3.3cm},clip]{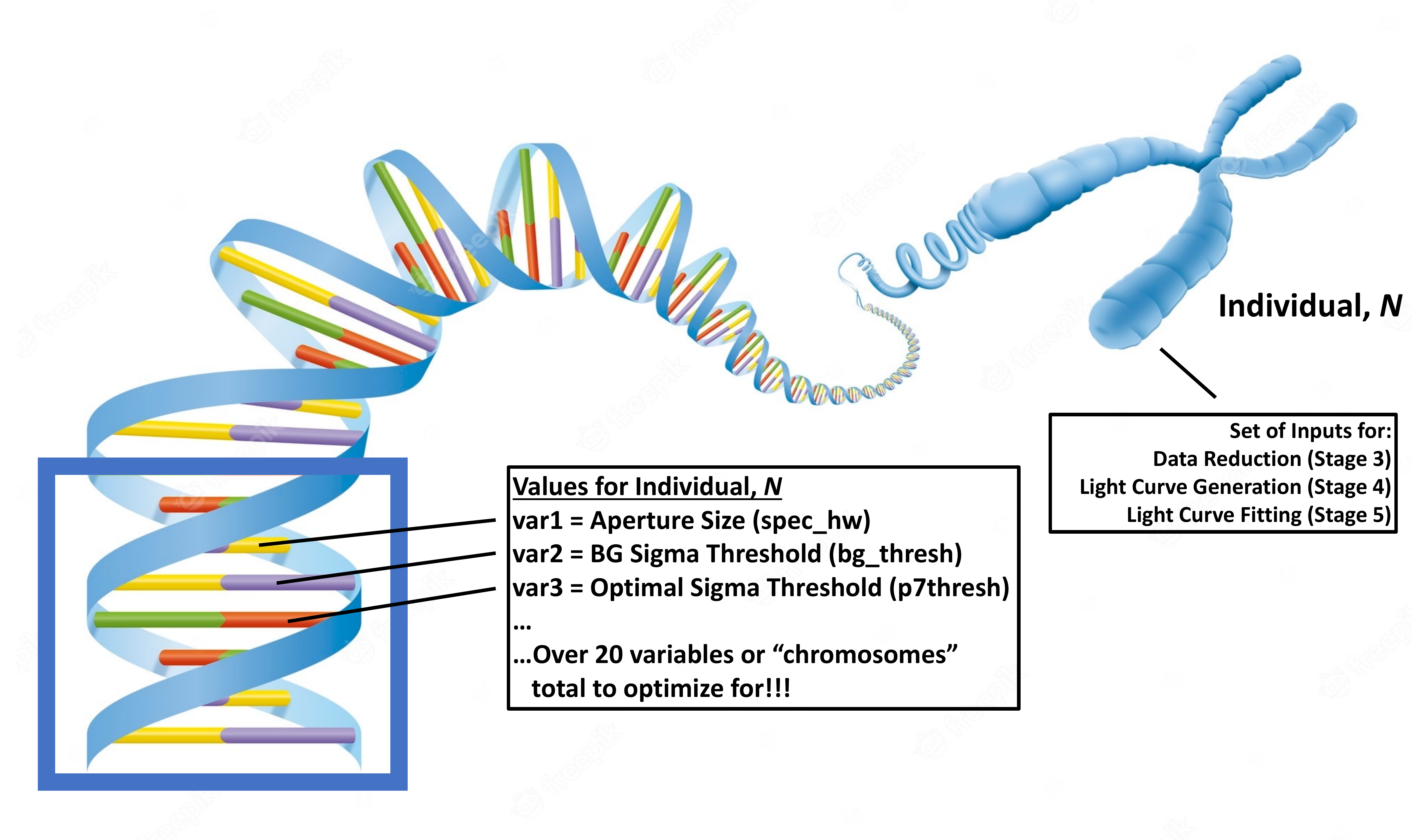}
    \end{minipage}
    \vspace{0cm} 
    \begin{minipage}{\textwidth}
        \centering
        \includegraphics[width=0.75\linewidth, trim={1.5cm 0cm 2cm 0cm},clip]{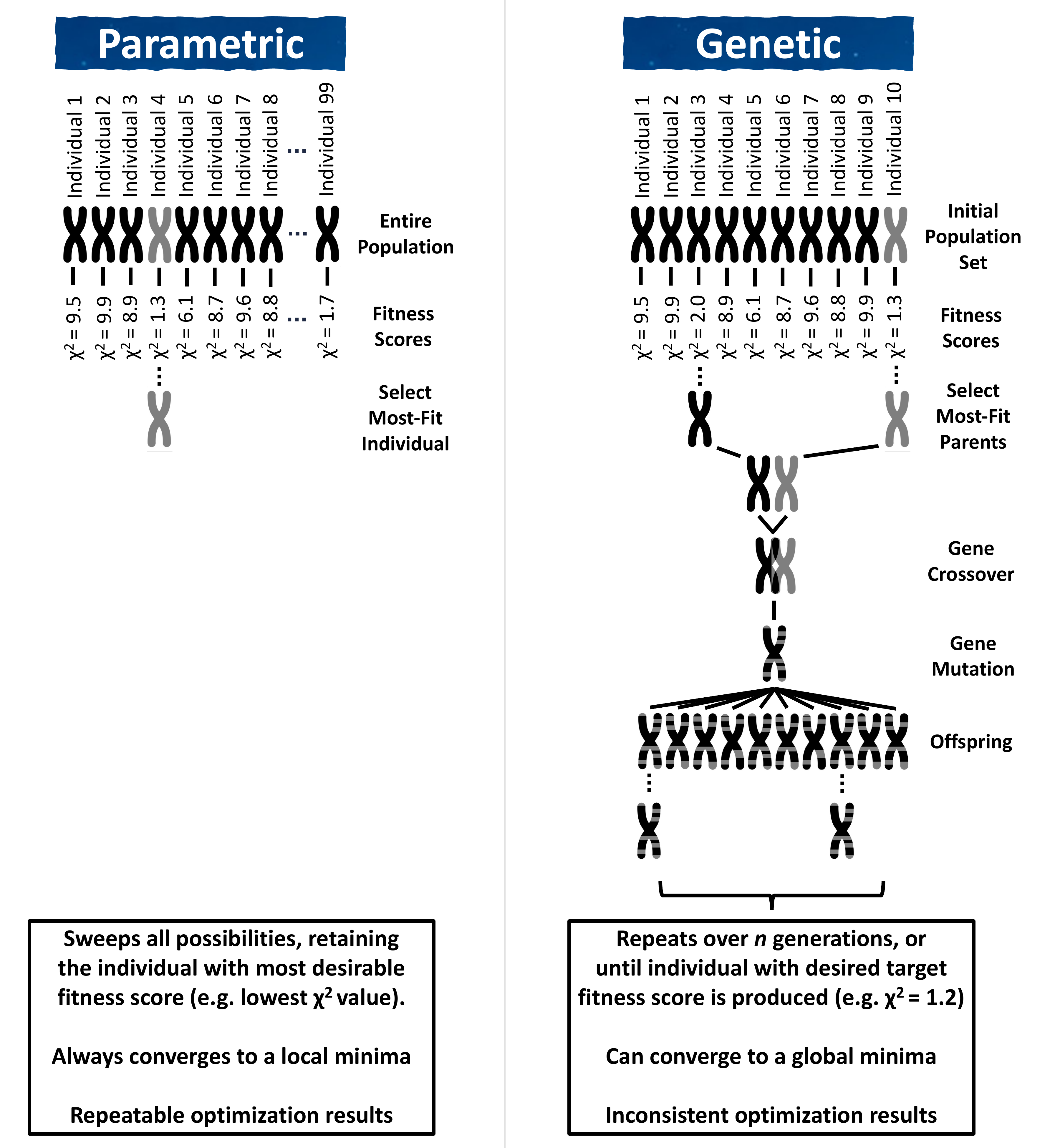}
            \vspace{0.5cm}
            \caption{\textit{Top} - An illustration of how each individual is "sequenced" with an assortment of variable-value pairs relative to transit processing. \textit{Bottom} - The process for how optimization via parametric sweeps and genetic algorithms operate. While parametric sweeps only optimize to local minima, their optimization is very repeatable, and thus are preferred to genetic optimizers for this application. }
    \label{fig:parametric_genetic}
    \end{minipage}
\end{figure*}

Parametric optimization (illustrated in \autoref{fig:parametric_genetic}) performs parametric sweeps across one or multiple variables in a sequential order, closely mimicking, but automating the manual optimization approach used by most astronomers' operation of \texttt{Eureka$!$} \citep{Bell2022}. This method tests all values within a specified range for each variable, selects the best choice, and then moves to the next variable(s) to be optimized. 

Genetic optimization (illustrated in \autoref{fig:parametric_genetic}), or optimization via genetic algorithms, tests random assortments of values within the specified bounds for the free parameters, and evaluates several variables simultaneously. These tests are run across a number of various assortments, with each run being referred to as an individual. The two best fit individuals are then selected as parents, their values are shared, and used to create a new generation of evolved / improved individuals compared to the population that existed in the first generation. This process is repeated for a number of generations until an individual with the demanded criteria (e.g. outstanding light curve quality, model-fit, etc.) is generated.

While genetic optimization offers the potential to converge to solutions within the global minimum, the randomness of the analysis means the number of generations required to converge to the solution may vary. Additionally, by ensuring an adequate population size is evaluated each generation, a genetic algorithm can effectively avoid converging to one of the first local minimum encountered during optimization, leading to more optimal solutions.  We have found that genetic optimization is most useful when the goal is to quickly explore a broad, multi-dimensional parameter space in search of the region containing the global minimum.

\begin{figure}[!]
    \centering
    \includegraphics[width=1.0\linewidth, trim={0.2cm 0cm 0cm 0cm},clip]{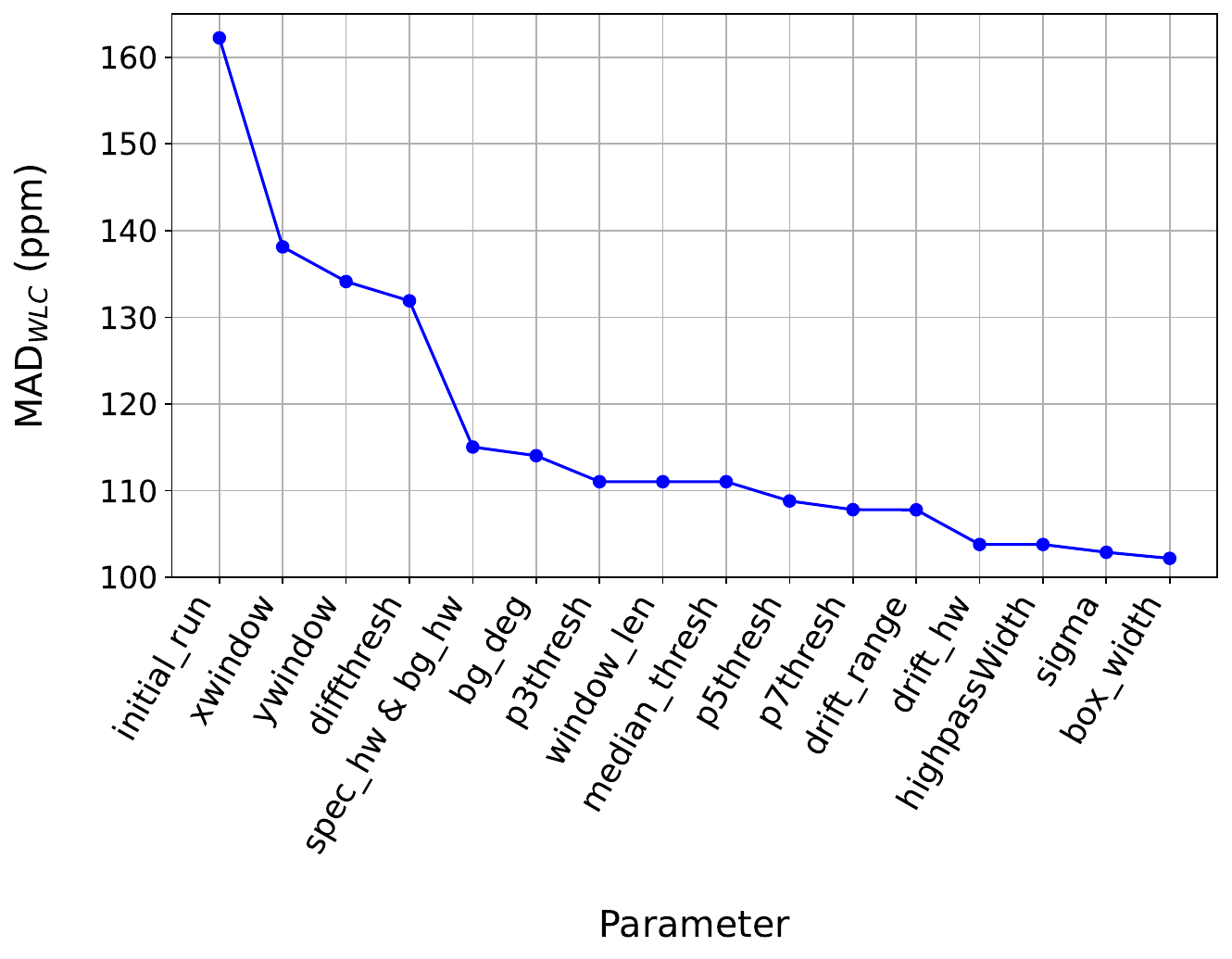} 
    \caption{An example output from \texttt{Eureka$!$}'s HST optimizer, demonstrating sequential optimization of the pipeline's parameters. For simplicity, the optimization shown here selects the best parameter values solely based on the MAD of the white light curve generated (MAD$_{WLC}$). As each parameter's optimal value is selected, it is stored and used in the optimization of the next parameter, yielding the best sequentially-optimized solution.}
    \label{fig:optimization_history}
\end{figure}

Parametric optimization on the other hand, while considered a more primitive form of automated, intelligent decision-making \citep{Charbonneau1995, Rahmat1999}, will always deliver an adequate solution that is at least in a local minimum. Most importantly, the optimization achieved via parametric sweeps is very repeatable. For this reason, parametric optimization is currently the preferred choice for \texttt{Eureka$!$} automation. An example of \texttt{Eureka$!$} parametrically optimizing for the best parameter values based on white light curve MAD value is shown in \autoref{fig:optimization_history}.

\begin{figure}[!]
    \begin{minipage}{0.481\textwidth}
        \centering
        \includegraphics[width=\linewidth, trim={2cm 0cm 1.5cm 0cm}, clip]{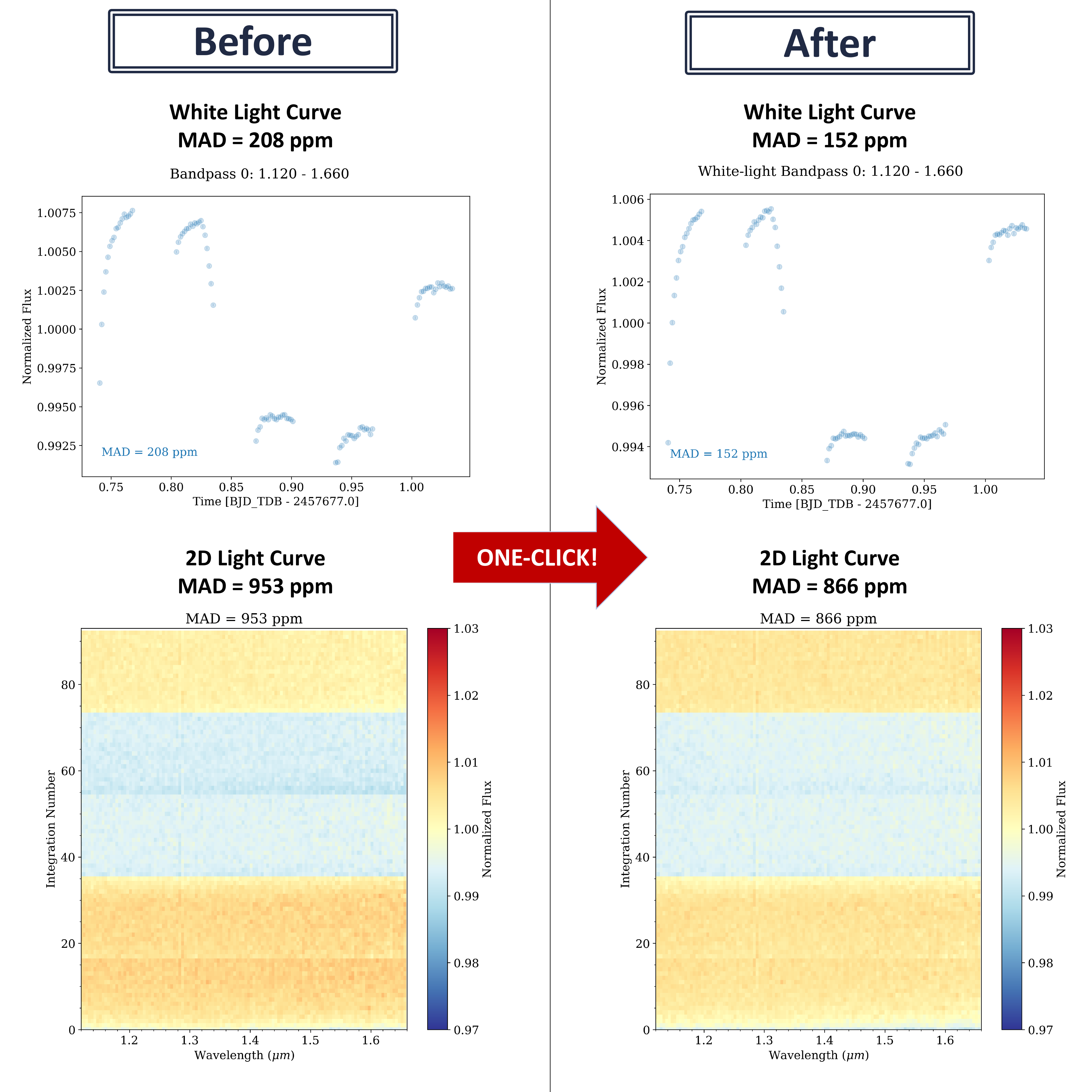}
    \end{minipage}%
    \vspace{0.1cm}

    \vfill
    \hrule
    \hrule
    \hrule
    \vfill
    
    \begin{minipage}{0.477\textwidth}
        \vspace{0.1cm}
        \includegraphics[width=\linewidth, trim={1.5cm 0cm 1.5cm 0cm}, clip]{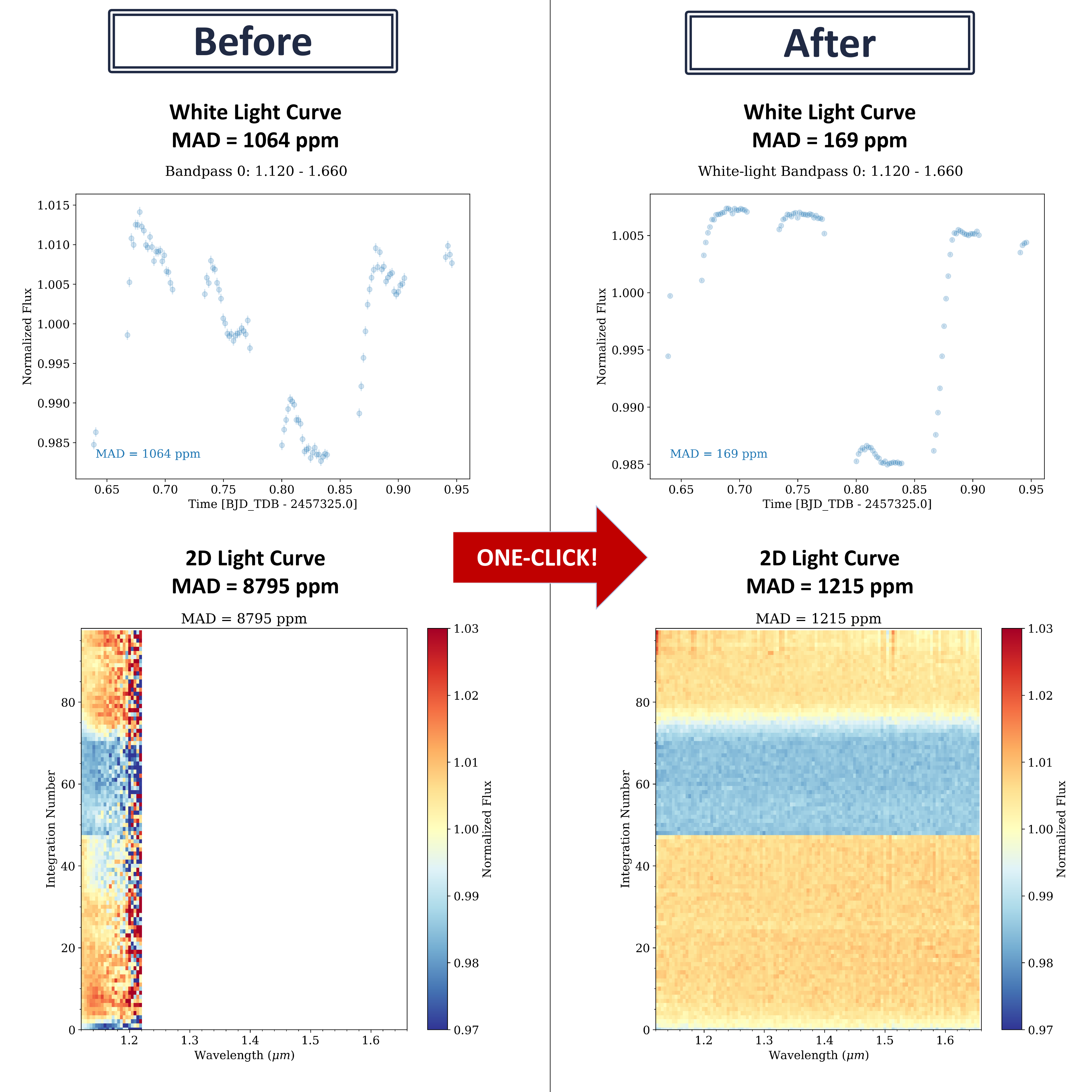}
    \end{minipage}
    \caption{Two separate case studies demonstrating proof-of-concept for the proposed AI-based processing, with HST WFC3 observations. \textit{Top} - White light and 2D light curves after initial processing and after AI-based processing. \textit{Bottom} - In this scenario, 
    \texttt{Eureka$!$}'s optimal estimates for the spectrum location were very poor, yielding an unusable data reduction. The AI corrected for this during its optimization, producing an outstandingly-improved result.}
    \label{fig:before_after}
\end{figure}

Compared to Bayesian retrieval techniques like Markov Chain Monte Carlo (MCMC) or nested sampling often used towards light curve fitting of exoplanets \citep{Kipping2014, Sharma2017, Bell2022}, the optimization methods introduced here differ in their focus and application. Rather than estimating the posterior distributions of physical model parameters, we automate the data reduction and analysis parameter selection within \texttt{Eureka$!$}. This can be viewed as a form of “hyper-retrieval", where the optimization operates on the user-defined configuration parameters that set up the retrieval process itself. By systematically exploring these parameters through parametric or genetic optimization, we aim to enhance the efficiency and repeatability of data analysis workflows, bridging the gap between manual tuning and fully automated optimization.

\subsection{Autonomous Data Reduction of HST Transits}
We have successfully developed a prototype for AI-based transit processing of HST observations. The prototype successfully demonstrates automated optimization for the data reduction and light curve generation stages of \texttt{Eureka$!$}, as shown in \autoref{fig:before_after}. For a detailed list of the parameters optimized using the optimization algorithm presented here, please refer to \autoref{table:optimized_parameters}. 

In this figure, we demonstrate the effectiveness of the AI-based optimizer, reducing what typically takes human staff days to accomplish, to a single click. Two cases are shown, highlighting the robustness of the AI-based optimizer: (1) A typical optimization, improving the initial data reduction by a decent amount; and (2) an outlier case with a very poor initial data reduction. In  the latter, initial guesses for the spectrum location barely contained any of the spectrum, causing a poor white light curve and unusable spectral data. While these were still used as initial values to kick-off the processing, the AI optimized the spectrum extraction dimensions as part of its programming, and corrected for this.



\section{Transmission Spectroscopy \\of Jovians and Sub-Neptunes}
\label{sec:water_findings}
Using the prototype AI-based HST optimizer, we have led a homogenous data reduction of 20 exoplanets (50 visits total) meeting preliminary quality-control requirements: (1) Less than $0.2$ pixels of peak-to-peak drift in the spectral direction of the detected image \citep{Stevenson2019}, and (2) an error-free white light curve containing a transit. Analyzing observations meeting these criteria from the \href{https://www.stsci.edu/~WFC3/trexolists/trexolists.html}{TrExoLiSTS} WFC3 transit database \citep{Nikolov2022}, target exoplanets observed with the WFC3 G141 mode (listed in \autoref{table:optimization_results}) were selected. The detailed list of all observations processed as part of this work are provided in a machine-readable format, available in the online-version of this article. The results of these automatically-optimized measurements are shown in \autoref{table:optimization_results}. 

The optimizer was programmed to evaluate the best outputs from \texttt{Eureka$!$} based on a fitness score of $f = 0.3*MAD_{white} + 1.0*MAD_{spec} + 0.0*{\chi^2}_{white}$. Here, the MAD values of light curve data are used to guide the optimizer towards yielding less scattered and outlier-ridden data. The equation for the fitness score is intended to produce the most beneficial white light curves and spectroscopic light curves are selected for generating accurate spectral data. The scalar values in front of each variable denote the weighting for each factor, where in this example, the $\chi^2$ fit of the transit model to the light curve is ignored. This weighting was determined to produce consistent, desirable results for parametrically optimizing transit measurements with \texttt{Eureka$!$}. 

The fitness score presented here was selected via rigorous testing, determining the $MAD_{spec}$ was the most critical component to yielding the most desirable transmission spectra. $MAD_{white}$ was determined to be less important in producing high-quality spectra, however an iterative process determined a weighting of $0.3$ was necessary for ensuring the white light curve was adequate for consistently producing desirable spectra.
Finally, visual-inspection and manual orbit-fitting were used to guarantee an adequate model fit to the light curves generated. Thus, a weighting of $0$ is applied to the $\chi^2$ fit of the white light curve to convey the omittance of light-curve fitting from the optimization used in this work. Using these algorithms to automate and optimize data reduction, in conjunction with manually-fitting light curves, high-quality transmission spectra are generated rapidly in a consistent, trusted manner.

On average, we observed an improvement of 30\% among white light curve MAD values, as well as a 36\% improvement to spectroscopic MAD values. We found a positive correlation of $r = 0.56$ between the pointing drift in the X-direction of the detector (the direction of the spectrum) and the ${\chi^2}_{\text{spec, median}}$ value. This correlation between pointing drift and limited light-curve fitting for spectroscopic data has been noted before \citep{Stevenson2019}. 

We present case studies from these optimizations in \autoref{fig:before_after} to provide a detailed look at the operation of the HST data optimizer. The presented AI-based processor successfully demonstrates automated optimization for the data reduction and light curve generation stages of \texttt{Eureka$!$}. 

The first case shown shows a standard case, with typical observation conditions, and a well-reduced outcome. In contrast, optimization of the second, more error-prone observation (\textit{HAT-P-18 b}, Program 14099, Visit 4) in \autoref{fig:before_after} highlights the robustness of the AI-based algorithm towards automating data reduction and lightcurve generation of exoplanet transit observations with \texttt{Eureka$!$}.

\begin{figure*}[p]  
\vspace{-1.25cm}
\centering
\begin{subfigure}
    \centering
    \includegraphics[width=\scalefactor\subfigwidth]{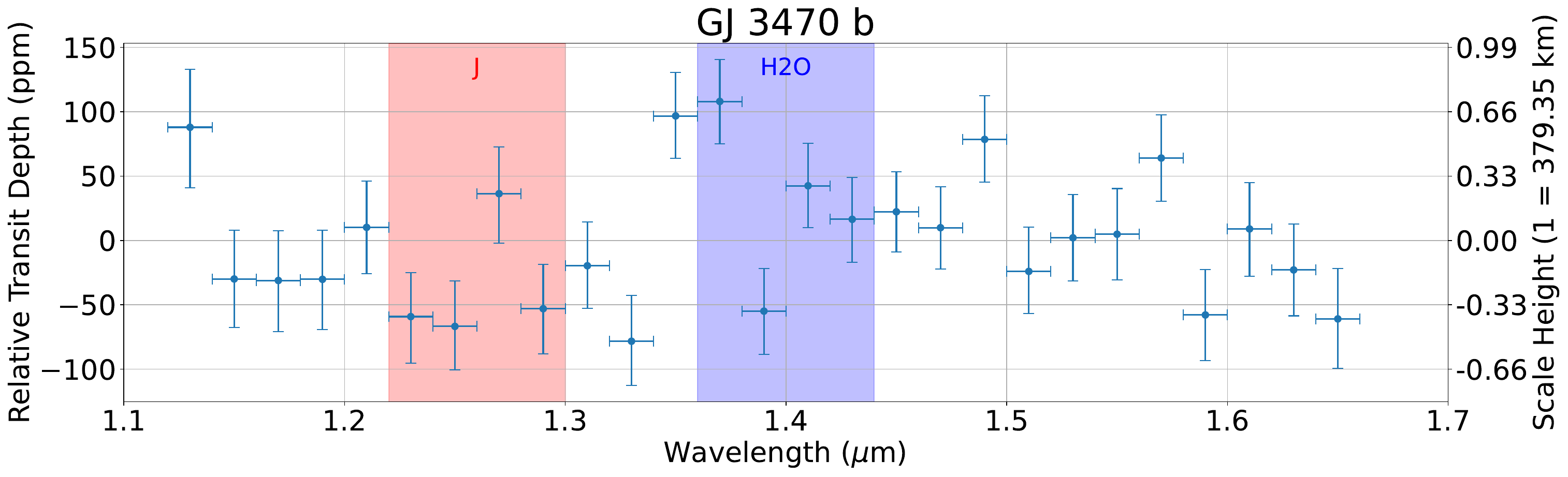} 
\end{subfigure}
\vspace{-0.25cm}
\begin{subfigure}
    \centering
    \includegraphics[width=\scalefactor\subfigwidth]{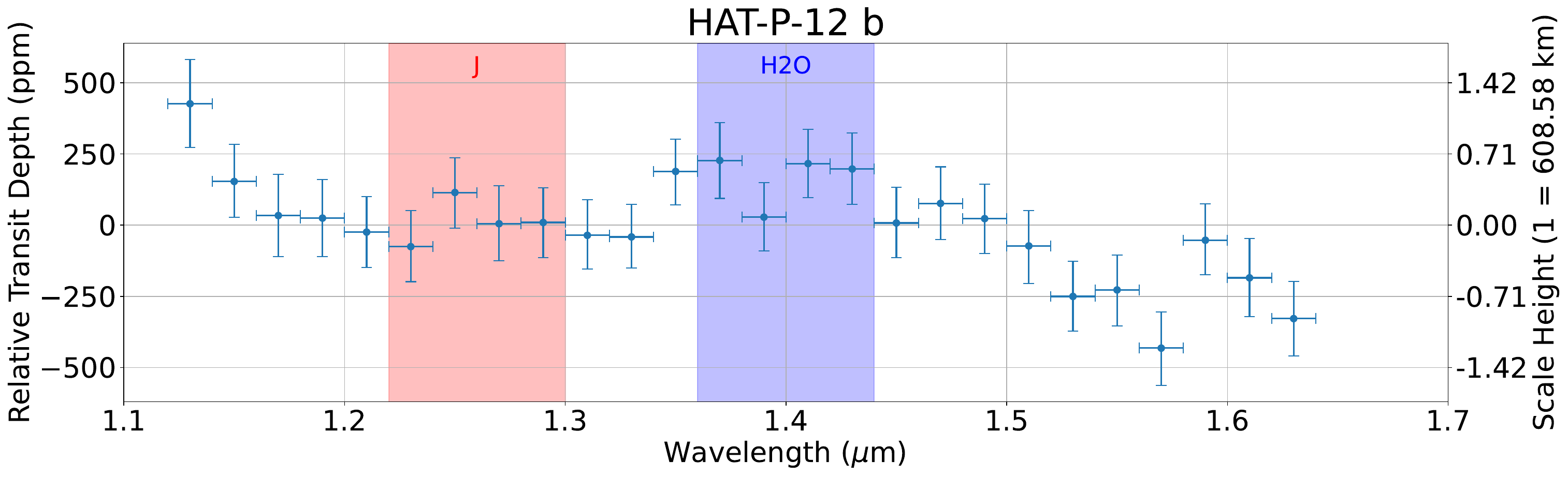} 
\end{subfigure}

\begin{subfigure}
    \centering
    \includegraphics[width=\scalefactor\subfigwidth]{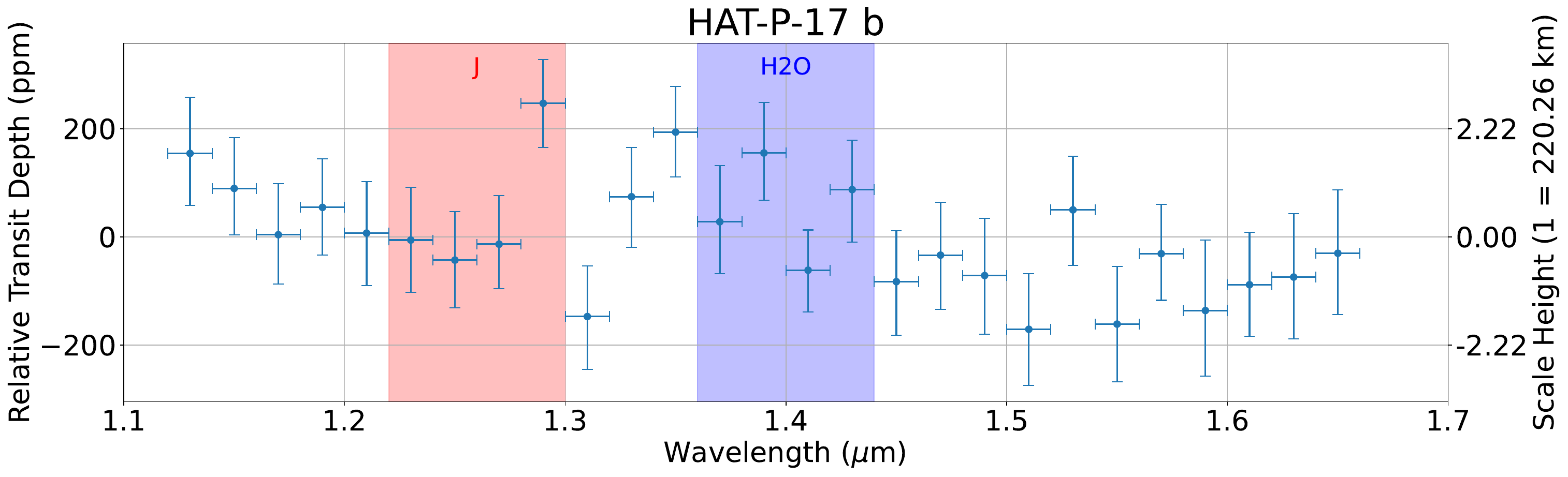}
\end{subfigure}
\vspace{-0.25cm}
\begin{subfigure}
    \centering
    \includegraphics[width=\scalefactor\subfigwidth]{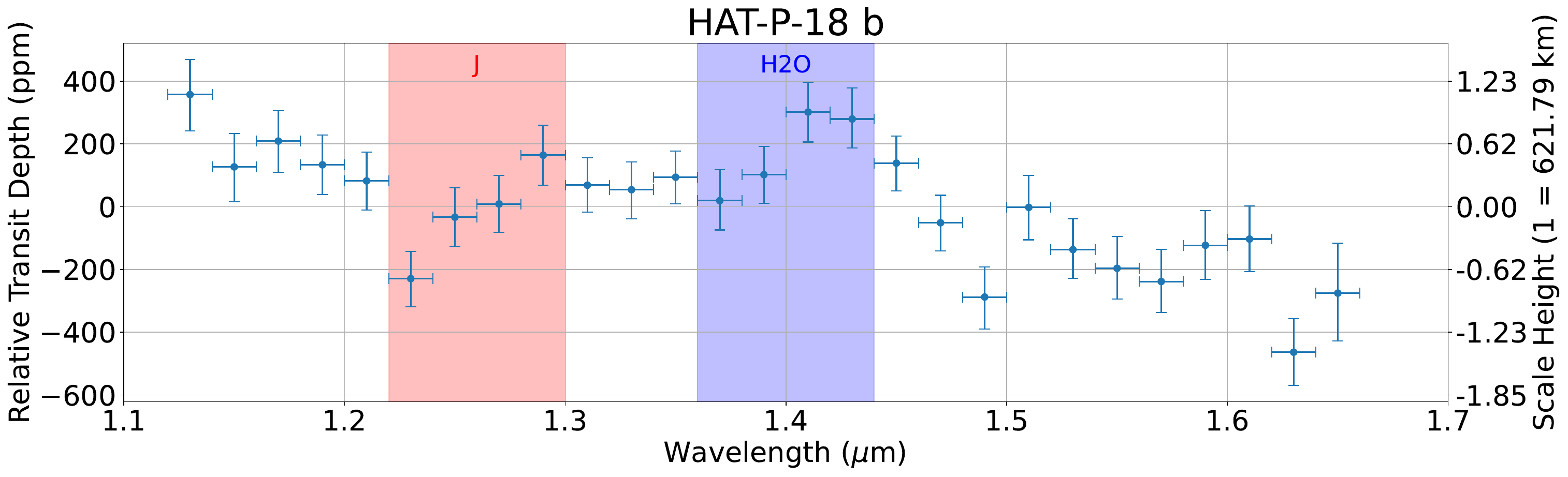}
\end{subfigure}

\begin{subfigure}
    \centering
    \includegraphics[width=\scalefactor\subfigwidth]{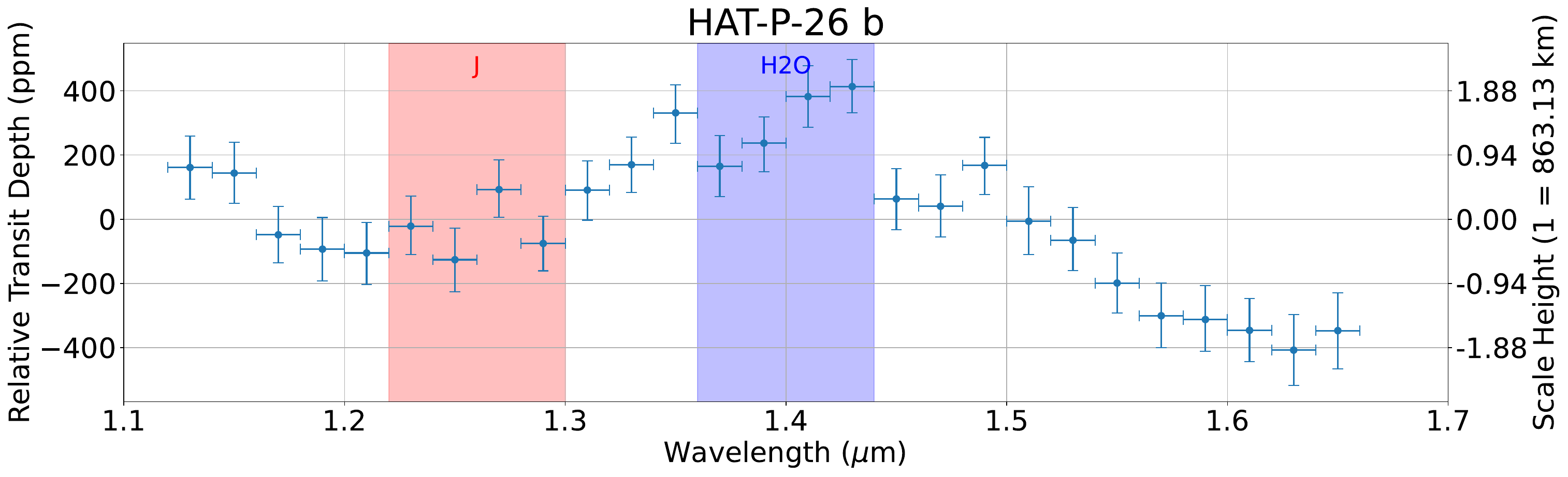}
\end{subfigure}
\vspace{-0.25cm}
\begin{subfigure}
    \centering
    \includegraphics[width=\scalefactor\subfigwidth]{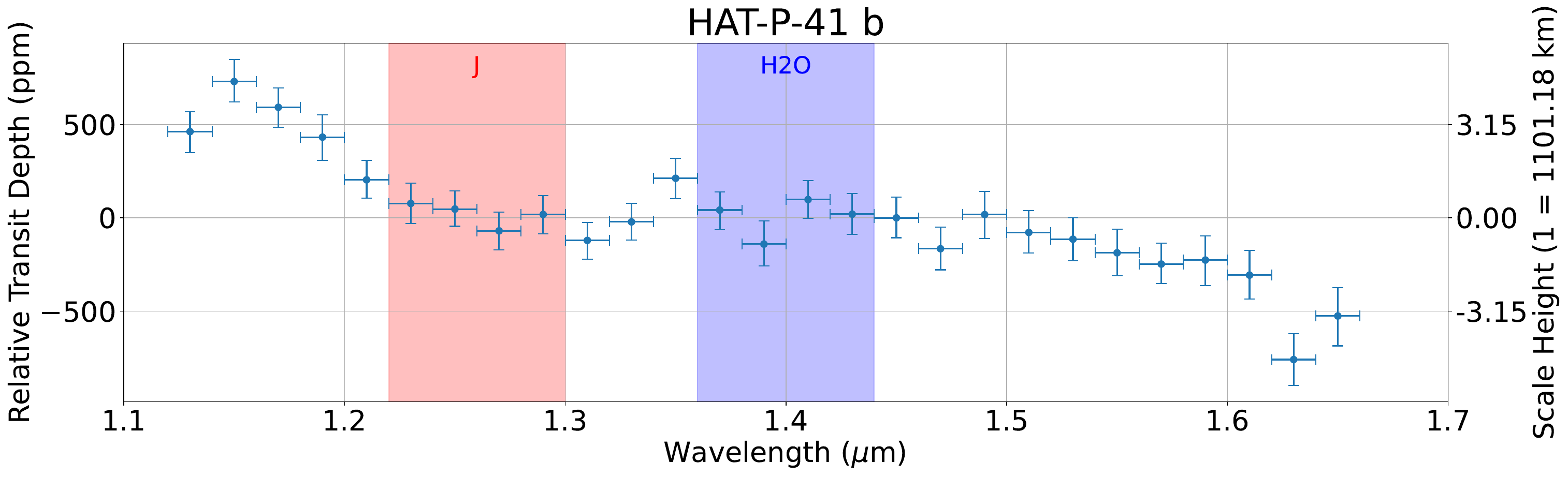}
\end{subfigure}

\begin{subfigure}
    \centering
    \includegraphics[width=\scalefactor\subfigwidth]{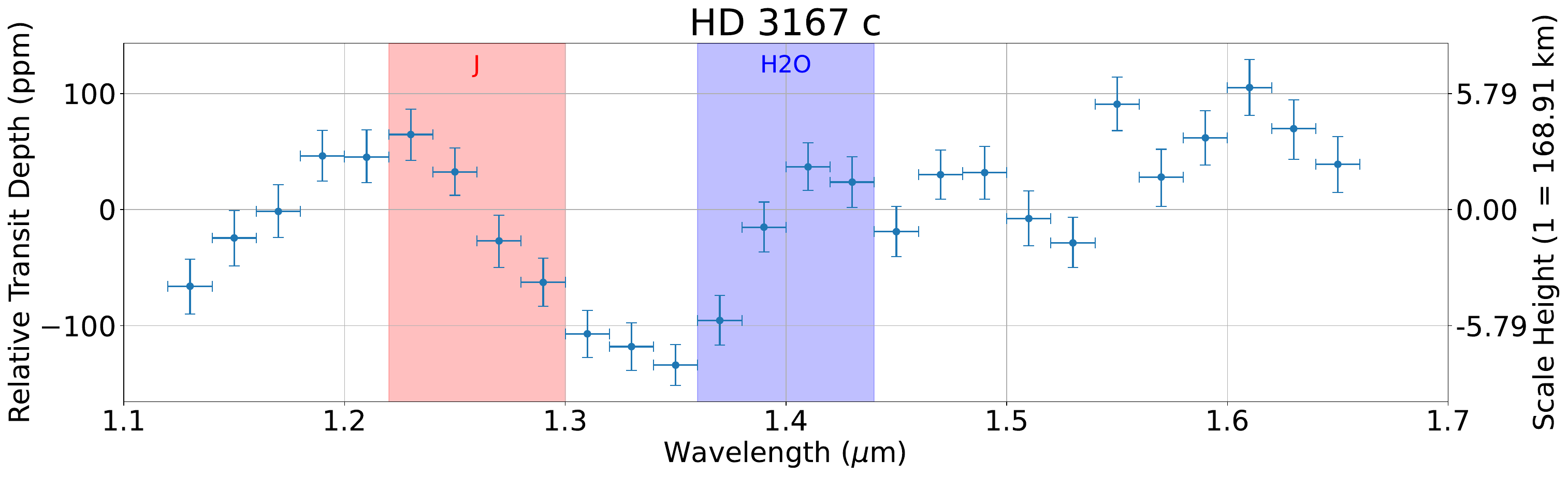}
\end{subfigure}
\vspace{-0.25cm}
\begin{subfigure}
    \centering
    \includegraphics[width=\scalefactor\subfigwidth]{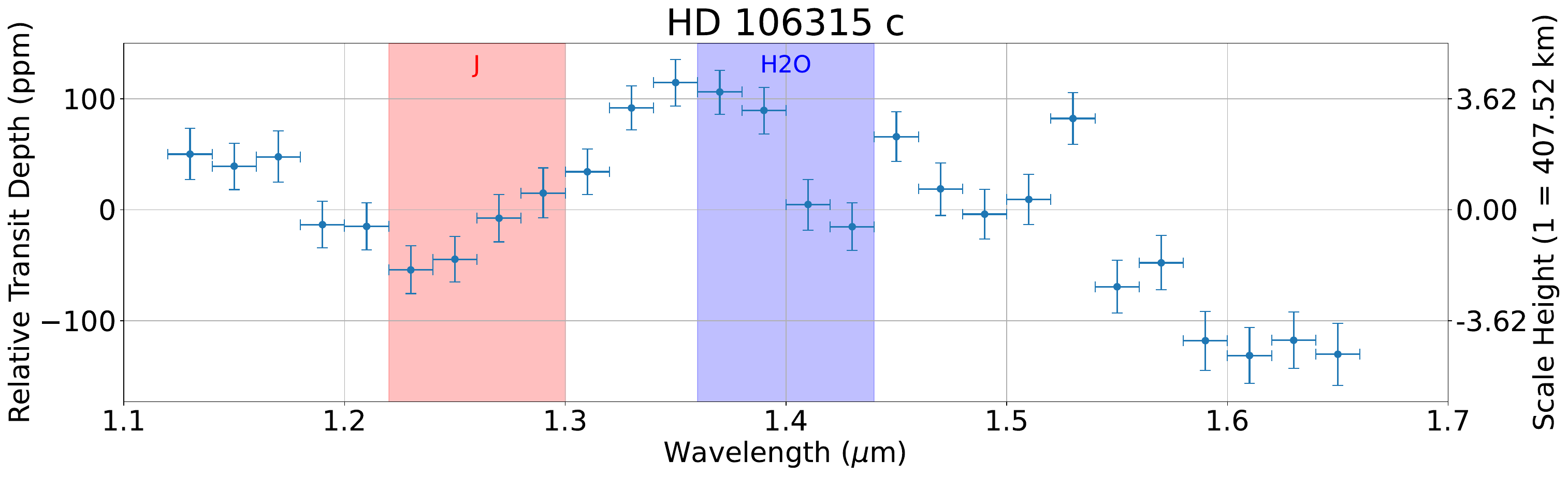}
\end{subfigure}

\begin{subfigure}
    \centering
    \includegraphics[width=\scalefactor\subfigwidth]{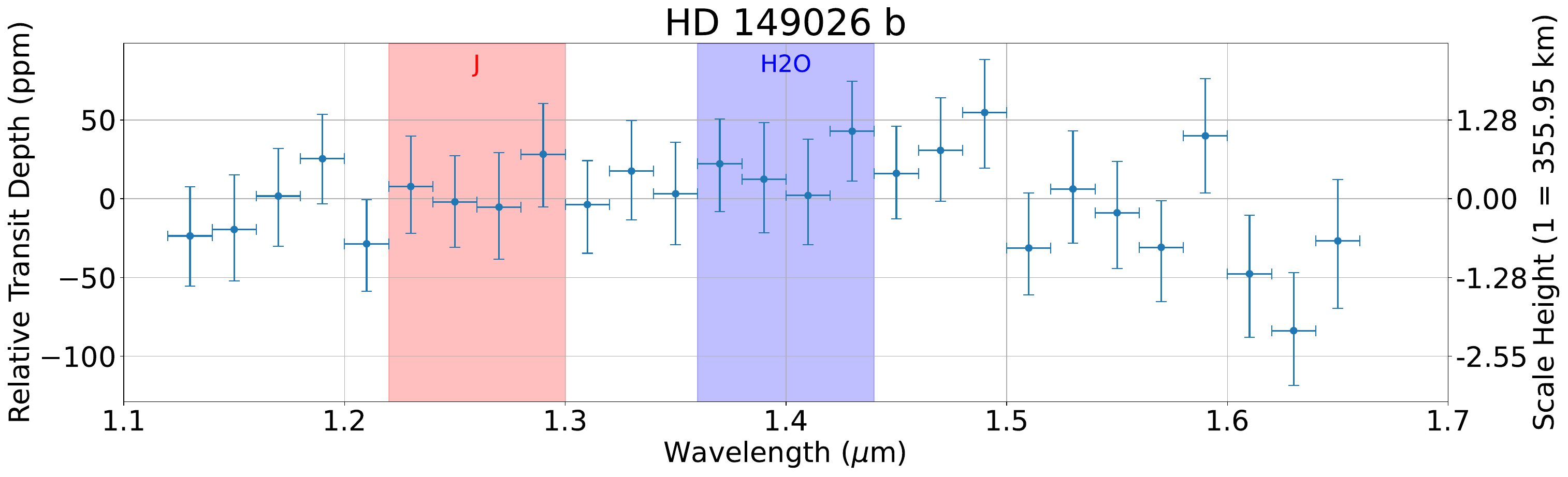}
\end{subfigure}
\vspace{-0.25cm}
\begin{subfigure}
    \centering
    \includegraphics[width=\scalefactor\subfigwidth]{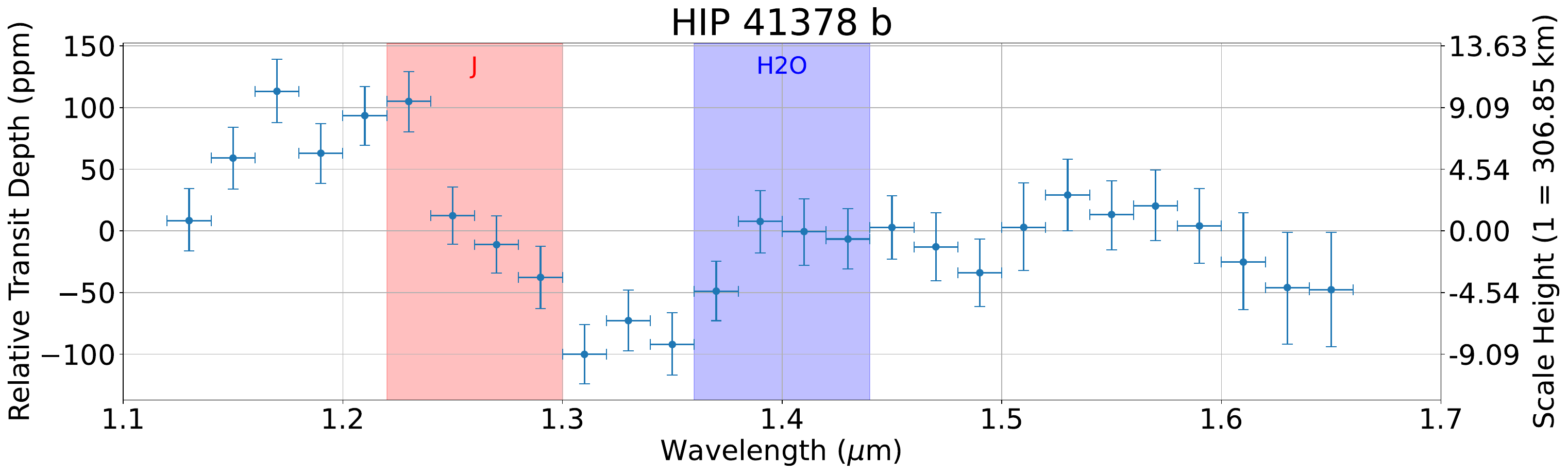}
\end{subfigure}

\begin{subfigure}
    \centering
    \includegraphics[width=\scalefactor\subfigwidth]{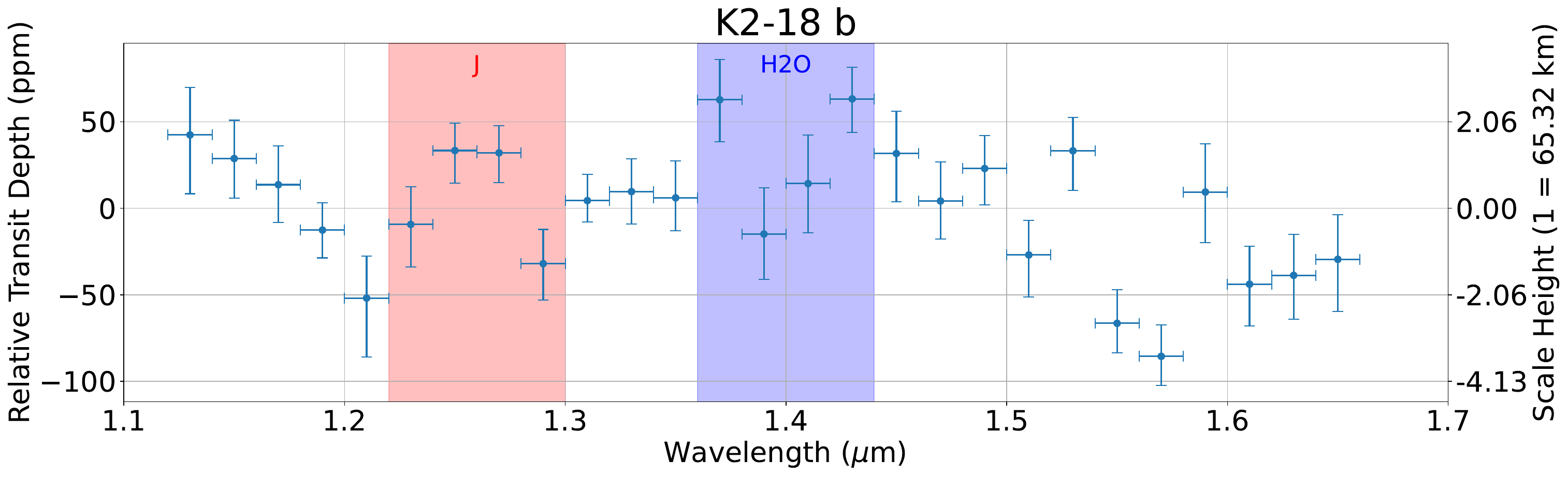}
\end{subfigure}
\vspace{-0.25cm}
\begin{subfigure}
    \centering
    \includegraphics[width=\scalefactor\subfigwidth]{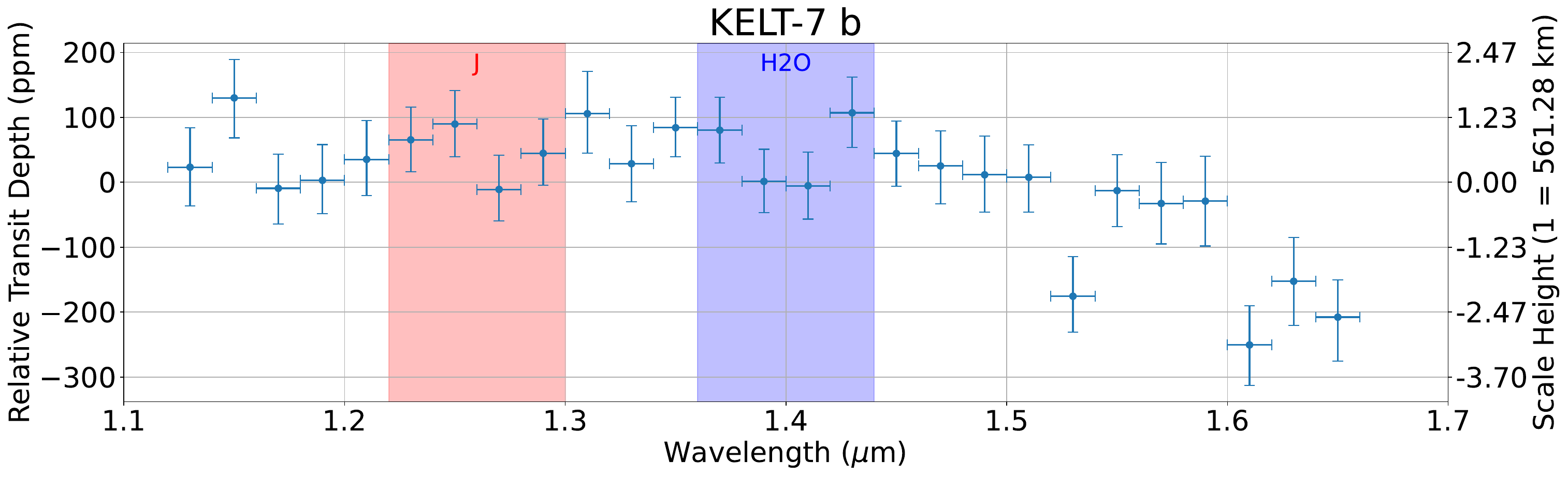}
\end{subfigure}

\begin{subfigure}
    \centering
    \includegraphics[width=\scalefactor\subfigwidth]{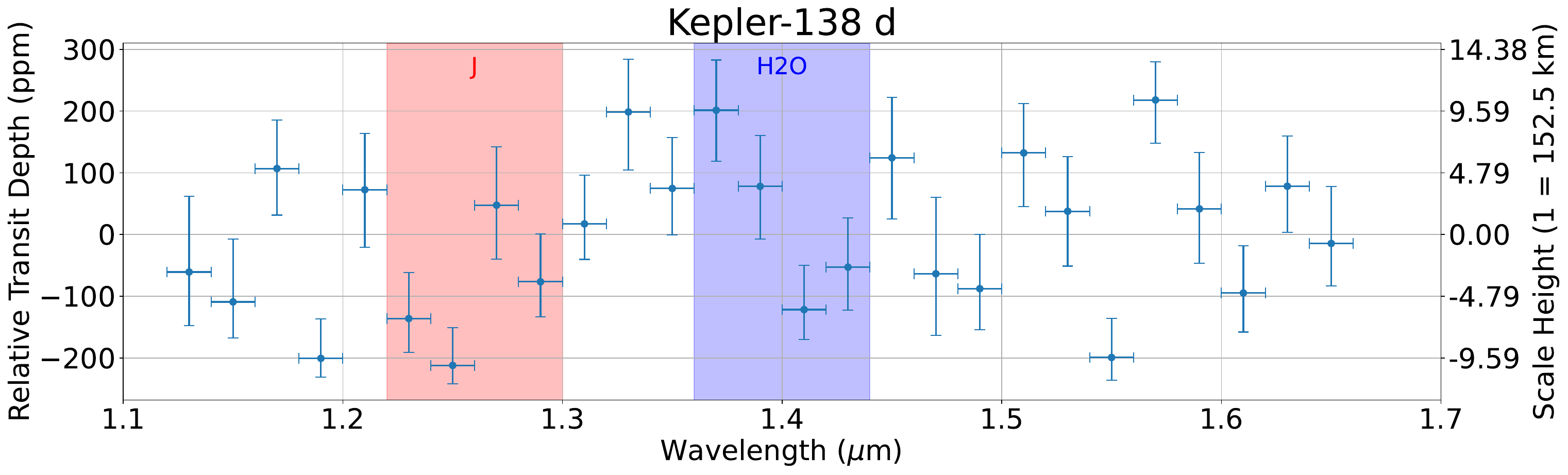}
\end{subfigure}
\vspace{-0.25cm}
\begin{subfigure}
    \centering
    \includegraphics[width=\scalefactor\subfigwidth]{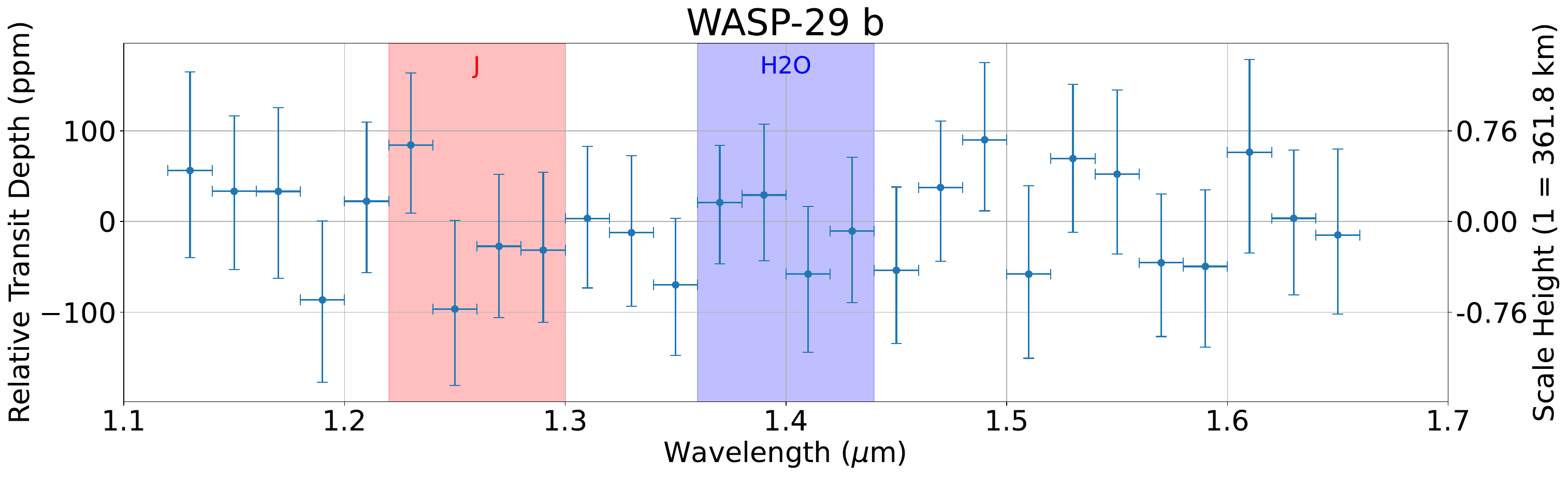}
\end{subfigure}

\begin{subfigure}
    \centering
    \includegraphics[width=\scalefactor\subfigwidth]{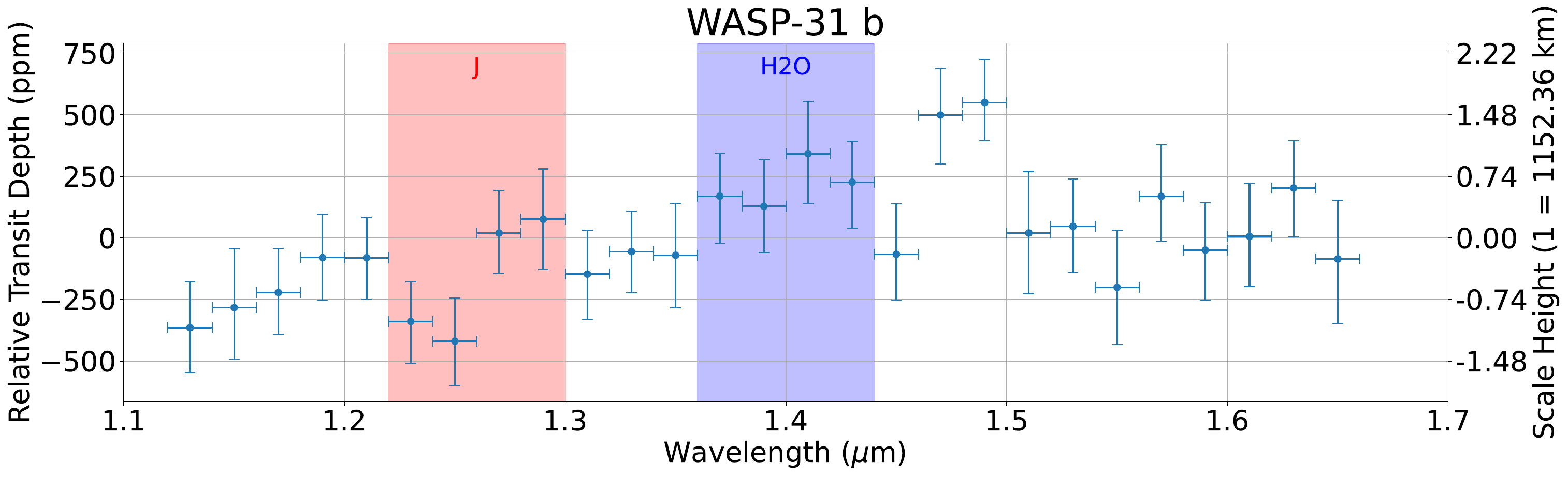}
\end{subfigure}
\vspace{-0.25cm}
\begin{subfigure}
    \centering
    \includegraphics[width=\scalefactor\subfigwidth]{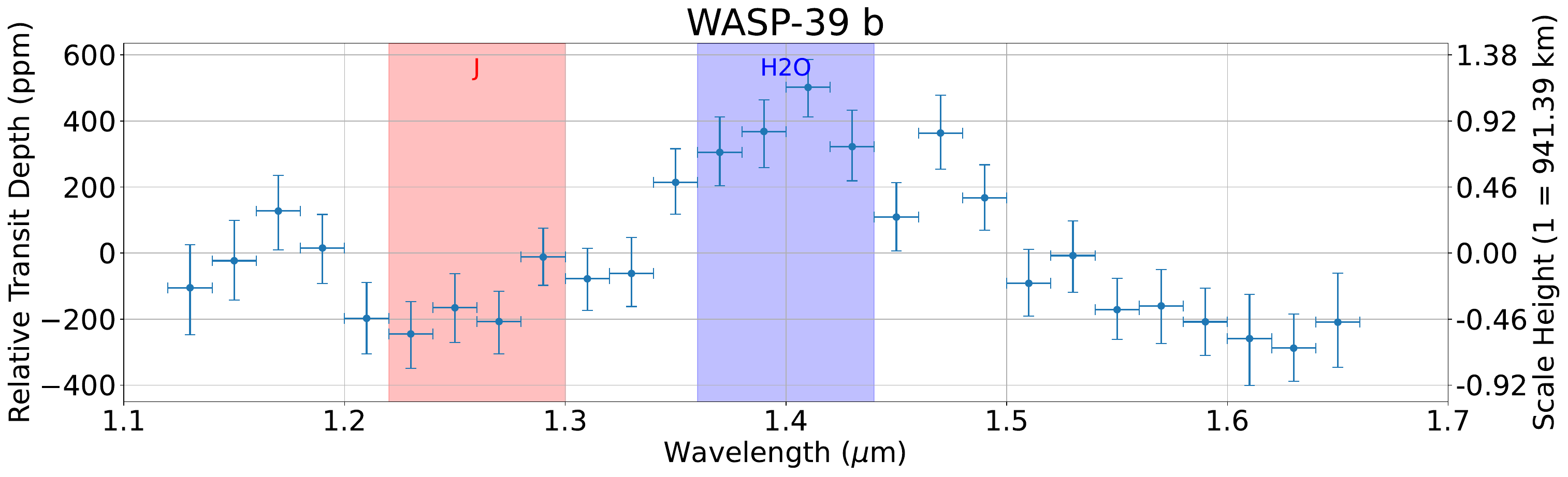}
\end{subfigure}

\begin{subfigure}
    \centering
    \includegraphics[width=\scalefactor\subfigwidth]{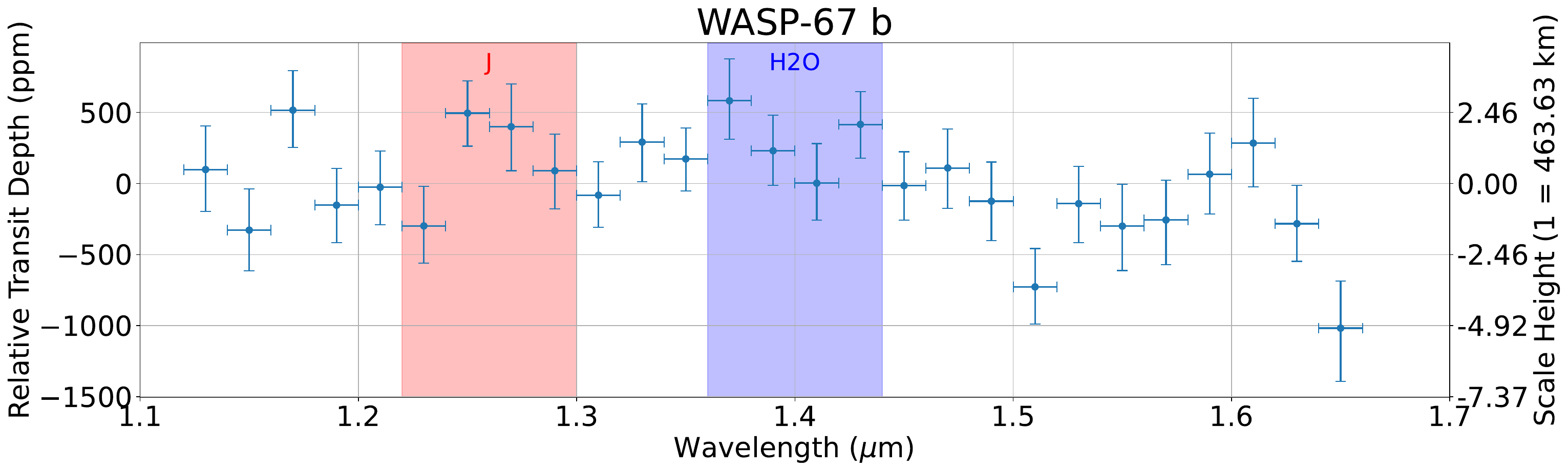}
\end{subfigure}
\vspace{-0.25cm}
\begin{subfigure}
    \centering
    \includegraphics[width=\scalefactor\subfigwidth]{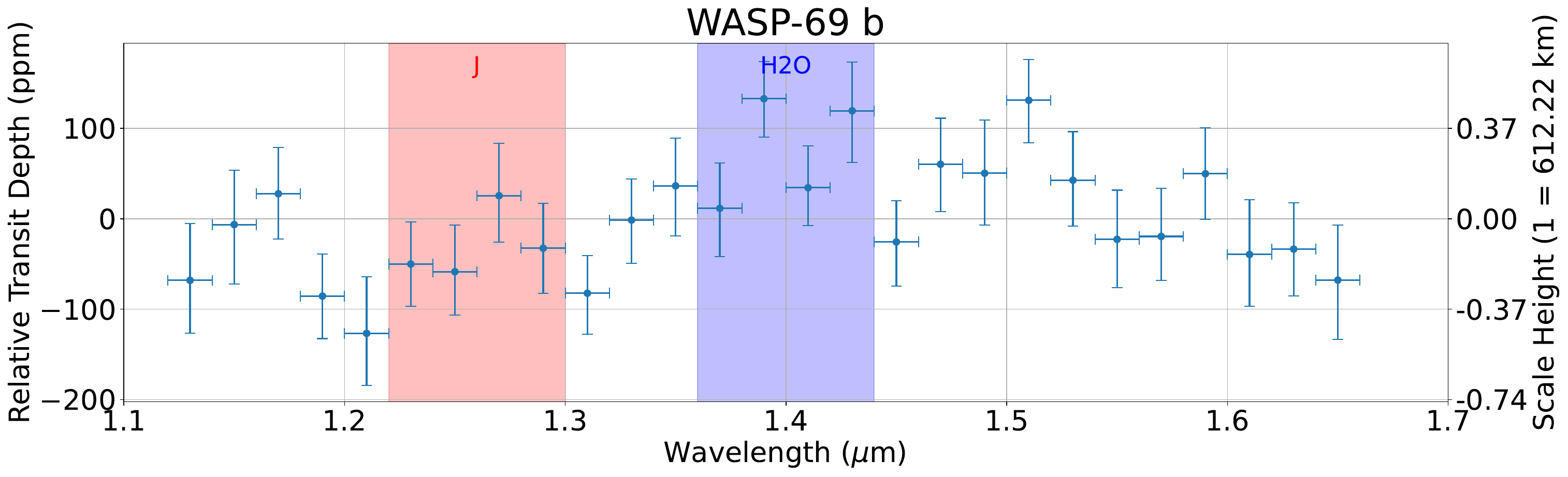}
\end{subfigure}

\begin{subfigure}
    \centering
    \includegraphics[width=\scalefactor\subfigwidth]{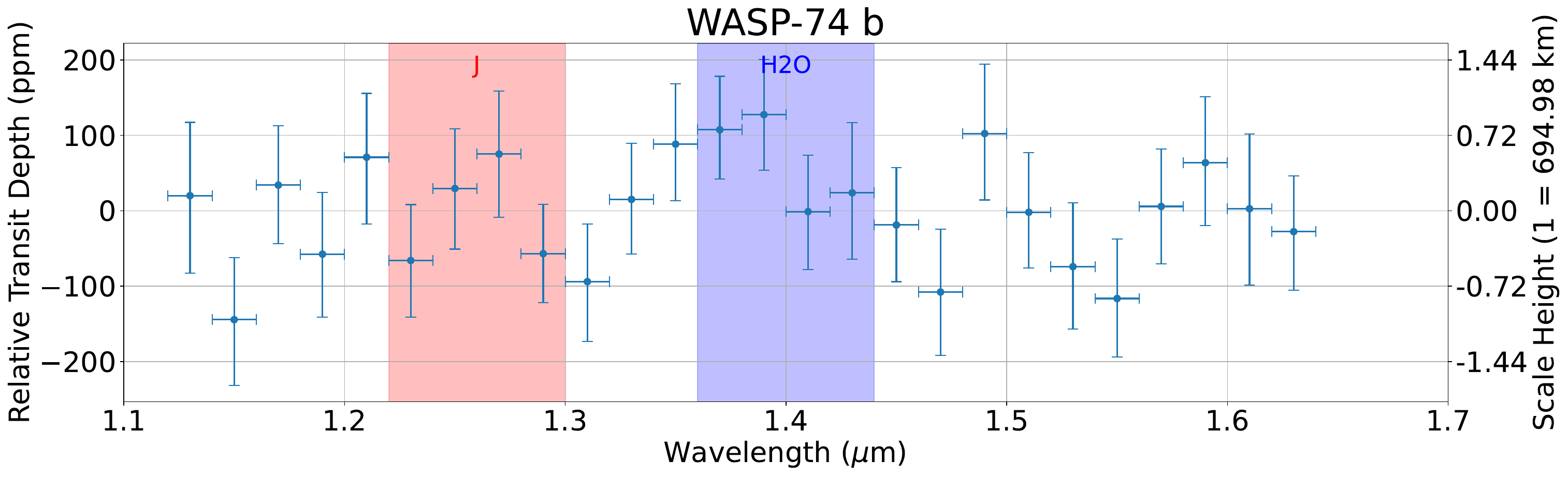}
\end{subfigure}
\vspace{-0.25cm}
\begin{subfigure}
    \centering
    \includegraphics[width=\scalefactor\subfigwidth]{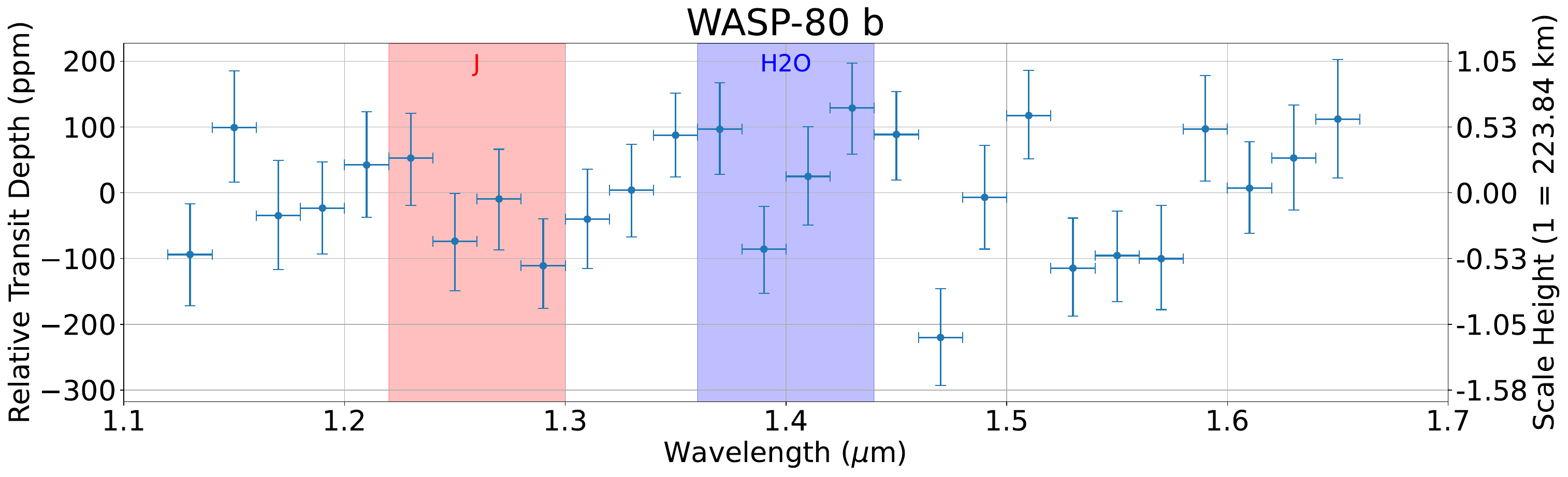}
\end{subfigure}

\caption{Transmission spectra of the 20 exoplanets analyzed as part of this survey.}
\label{fig:spectra}
\end{figure*}

Using this standardized, automated process for processing HST transits, we use the AI-based HST optimizer to generate transmission spectra for all exoplanet observations included in this work. The resulting transmission spectra of these exoplanets' atmospheres are shown in \autoref{fig:spectra}. 

\begin{table*}[!]
\centering
\rotatebox{90}{%
\begin{minipage}{\textheight}
\caption{Initial and optimized metrics for the AI-based optimization. A consistent improvement was observed across all white light curve MAD values, as well as the MAD values of all spectroscopic light curves for the datasets optimized in this survey, proving the efficacy of the optimizer. A mean improvement of 30\% was achieved among white light curve MAD values, as well as a 36\% improvement to spectroscopic MAD values. A moderately positive correlation of $r = 0.56$ was found between the pointing drift in the X-direction of the detector (the direction of the spectrum), and the ${\chi^2}_{\text{spec, median}}$ value, indicating a pointing limitation on how well the spectroscopic light curves are fitted \citep{Stevenson2019}.}
\label{table:optimization_results}
\vspace{1cm}
\hspace{-3cm}
\scalebox{1.15}{%
\resizebox{\textwidth}{!}{%
\begin{threeparttable}
\begin{tabular}{ccccccccc}
\textbf{Target}       & \textbf{Program ID} & \textbf{Visit} & \textbf{MAD$_{\text{white, initial}}$} & \textbf{MAD$_{\text{white, optimized}}$} & \textbf{MAD$_{\text{spec, initial}}$} & \textbf{MAD$_{\text{spec, optimized}}$} & \textbf{Drift$_X$ ($pk-pk$)} & \textbf{${\chi^2}_{\text{spec, median}}$} \\ \hline
GJ-3470 b             & 13665               & 24             & 185                        & 143                          & 577                         & 545                           & 0.14                      & 1.14                           \\
HAT-P-12 b            & 14260               & 16             & 384                        & 308                          & 1330                        & 1217                          & 0.24                      & 1.43                           \\
HAT-P-17 b            & 12956               & 2              & 162                        & 159                          & 713                         & 704                           & 0.27                      & 1.10                           \\
HAT-P-18 b            & 14099               & 4              & 1064                       & 169                          & 8795                        & 1215                          & 0.17                      & 1.13                           \\
HAT-P-26 b            & 14260               & 20             & 159                        & 133                          & 824                         & 822                           & 0.10                      & 1.14                           \\
HAT-P-41 b            & 14767               & 87             & 208                        & 152                          & 953                         & 866                           & 0.15                      & 1.11                           \\
HD-106315c            & 15333               & 14             & 208                        & 188                          & 430                         & 425                           & 0.32                      & 1.39                           \\
HD-149026 b           & 14260               & 9              & 93                         & 82                           & 318                         & 322                           & 0.08                      & 1.33                           \\
HD-3167c              & 15333               & 47             & 131                        & 114                          & 414                         & 374                           & 0.13                      & 2.04                           \\
HIP-41378 b           & 15333               & 56             & 250                        & 206                          & 421                         & 418                           & 0.40                      & 2.40                           \\
K2-18 b               & 13665               & 29             & 188                        & 170                          & 717                         & 706                           & -                         & -                              \\
KELT-7 b              & 14767               & 91             & 147                        & 144                          & 699                         & 664                           & -                         & -                              \\
Kepler-138 d          & 13665               & 1              & 243                        & 202                          & 904                         & 903                           & 0.15                      & 1.11                           \\
WASP-29 b             & 14260               & 15             & 149                        & 165                          & 644                         & 635                           & 0.09                      & 1.14                           \\
WASP-31 b             & 12473               & 25             & 199                        & 139                          & 1164                        & 1118                          & 0.11                      & 1.14                           \\
WASP-39 b             & 14260               & 10             & 300                        & 233                          & 1161                        & 1076                          & 0.15                      & 1.17                           \\
WASP-67 b             & 14260               & 7              & 277                        & 220                          & 1303                        & 1169                          & 0.15                      & 1.16                           \\
WASP-69 b             & 14260               & 14             & 178                        & 171                          & 496                         & 486                           & 0.15                      & 1.12                           \\
WASP-74 b             & 14767               & 79             & 138                        & 92                           & 560                         & 519                           & 0.08                      & 1.15                           \\
WASP-80 b             & 14260               & 5              & 225                        & 208                          & 638                         & 624                           & 0.11                      & 1.35                           \\
\hline
Avg.                  &                     &                & 244                        & 170                          & 1153                        & 740                           &                           & 1.31                           \\
\hline
Avg. Improvement      &                     &                &                            & 74                           &                             & 413                           &                           &                                \\
\hline
Avg. Improvement (\%) &                     &                &                            & \textbf{-30\%}               &                             & \textbf{-36\%}                &                           &                               
\end{tabular}%
\begin{tablenotes}\footnotesize
\item[\hspace{2cm}*] Initial conditions already use optimal estimates for extraction box window, spectrum aperture \& background sizes.
\item[\hspace{2cm}*] Results shown in this study have been optimized to fitness score of $f = 0.3*MAD_{white} + 1.0*MAD_{spec} + 0.0*{\chi^2}_{white}$.
\item[\hspace{2cm}*] For targets with multiple transits, the first visit's optimized values are used for all visits. This action was taken as a time-saving practice; however, optimization of each transit is recommended.
\item[\hspace{2cm}*] Light curves for K2-18 b and KELT-7 b were fitted using simultaneous fitting of spectroscopic light curves, due to errors encountered with Eureka!'s white light curve fitting for those visits.
\end{tablenotes}
\end{threeparttable}
}
}
\end{minipage}}
\end{table*}

Each of the 20 transmission spectra shown in \autoref{fig:spectra} has two highlighted wavelength bands: 1) A photometric J-band from $1.22-1.30$ $\mu m$ and 2) a \ce{H2O} band from $1.36-1.44$ $\mu m$ \citep{Stevenson2016}. Using these bands as references, we quantify the amplitude of the water feature in a consistent and uniform manner. With these relative measurements, we perform a population-level study in \autoref{sec:comparative} to identify trends in aerosol formation across the variety of exoplanet-types studied as part of this effort.

\section{Comparative Exoplanetology}
\label{sec:comparative}

\begin{figure*}[!]
    \vspace{-0.75cm} 
    \centering
    \begin{minipage}{\textwidth}
        \hspace{0.3cm}
        \includegraphics[width=1.02\linewidth, trim={0cm 0cm 0cm 0cm},clip]{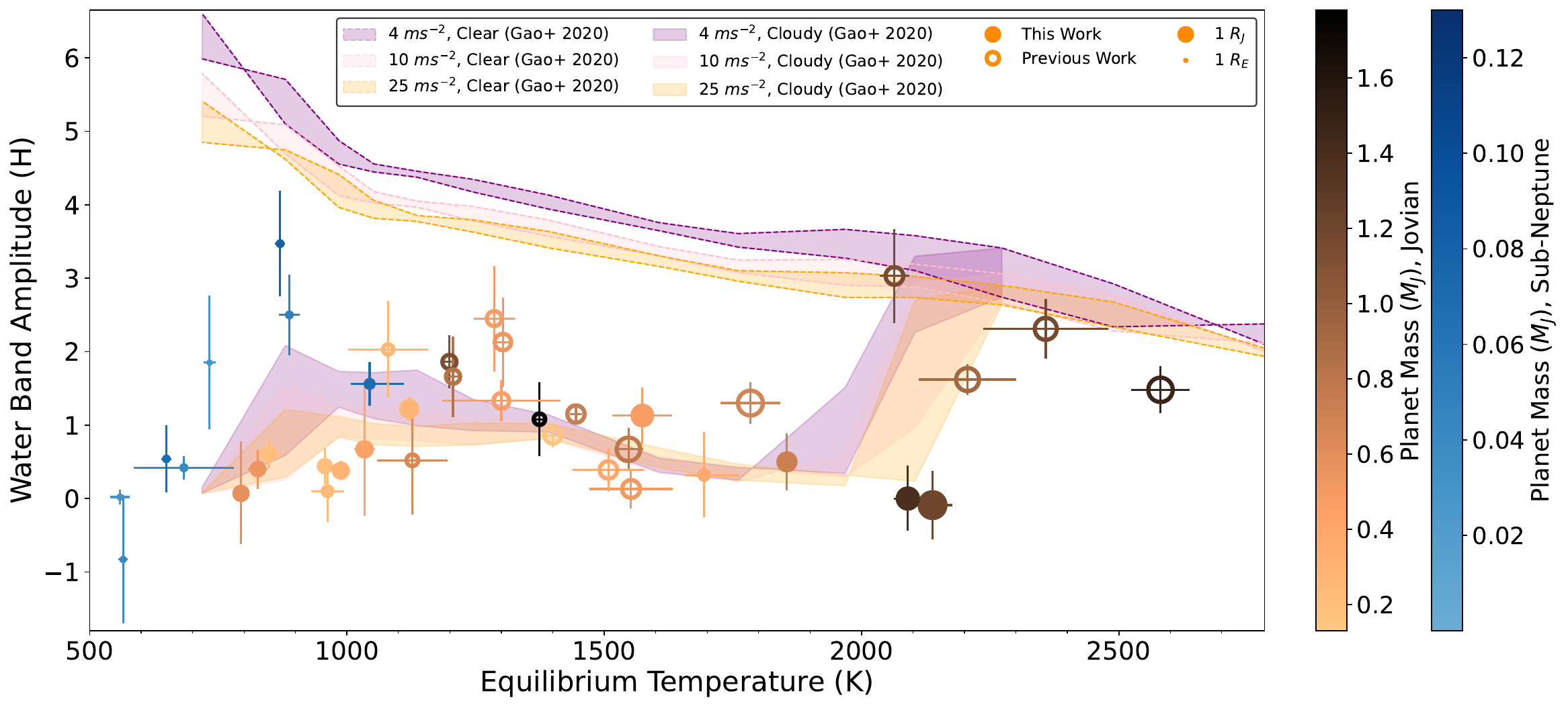}
    \end{minipage}
    
    \begin{minipage}{\textwidth}
        \hspace{-0.12cm}
        \includegraphics[width=0.978\linewidth, trim={0cm 0cm 0cm 0cm},clip]{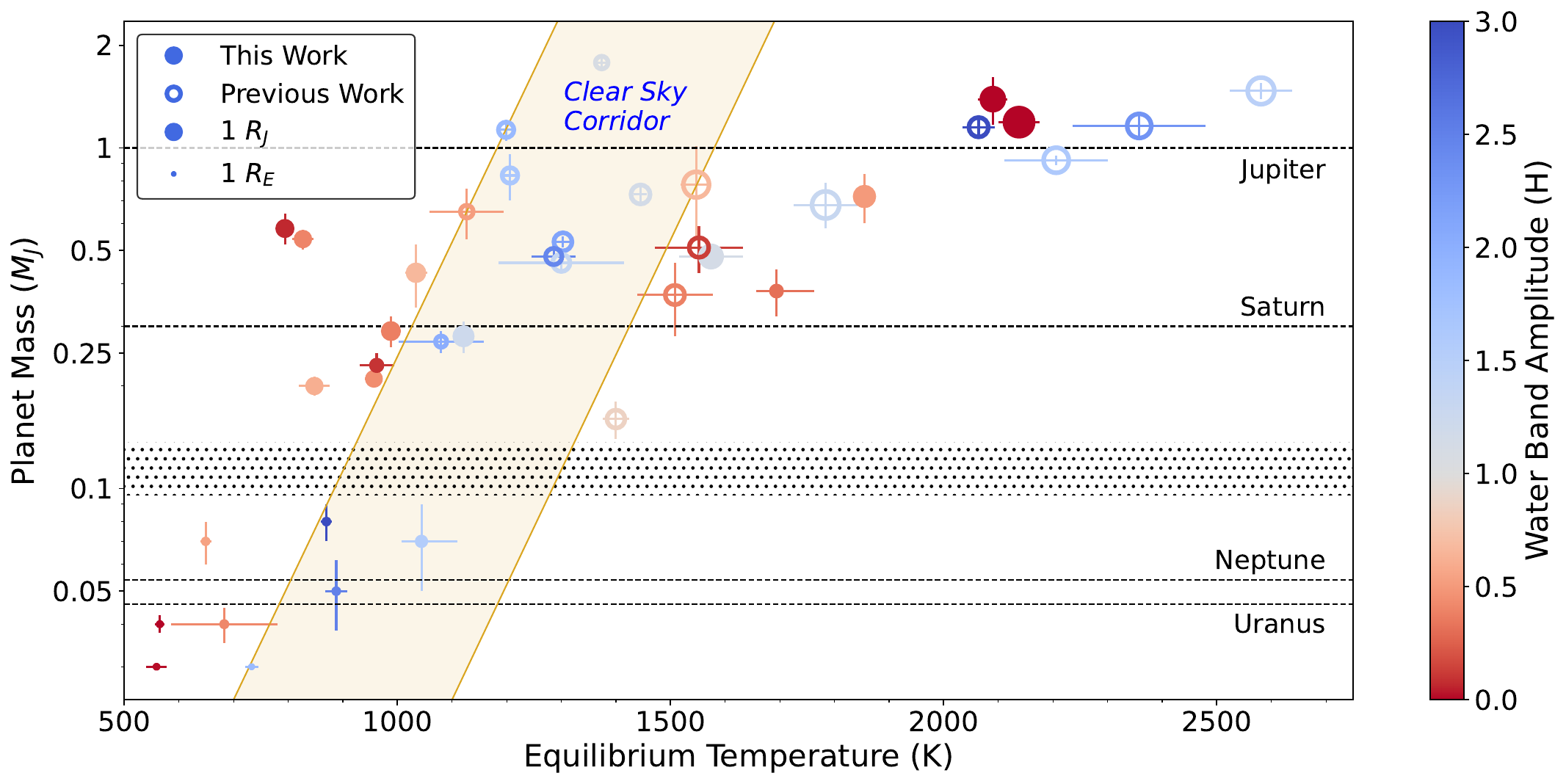}
        \caption{Aerosol-formation trends observed between regimes of Jovian and sub-Neptune class exoplanets. \textit{Top} - Jovian and sub-Neptunes analyzed in this and previous works, compared against notable published models for Jovian-aerosol-formation \citep{Gao2020}. \textit{Bottom} - Jovian and sub-Neptunes analyzed in this and previous works. Jovian exoplanets are shown above the dotted boundary region ($\sim$ $0.1$ $-$ $0.13$ $M_J$), and sub-Neptunes are shown below. A novel, separate pattern for \ce{H2O} aerosol formation among the sub-Neptune regime is clear. Furthermore, a newly-found corridor yielding a prominent increase in measured water-band amplitudes is evident.  This region presents favorable, clear sky conditions towards identifying atmospheric chemical species across exoplanet regimes, and is defined by a yellow-shaded region. The condensation point for forsterite (\ce{Mg2SiO4}) and aluminum oxide (\ce{Al2O3}), located at $\sim 2100$ K, causes the anomalous decrease in water feature amplitude observed in some of the hot Jupiters \citep{Gao2020}.}
        \label{fig:plots_Jovian_SubNeptune}
    \end{minipage}
\end{figure*}

The population of exoplanets surveyed in this paper (shown in \autoref{fig:plots_Jovian_SubNeptune}) contains information from 20 homogeneous exoplanet transmission spectra produced in this work, and \ce{H2O} band measurements of 22 other exoplanets from notable previous works \citep{Stevenson2016, Fu2017, Gao2020, Spake2021}.

\begin{table*}[!]
\caption{Measurements of the relative water band amplitude ($H_2O - J$) and relevant parameters for the exoplanets included in this survey and previous works. Six of the planets were repeated between this survey and notable previous works \citep{Stevenson2016, Gao2020, Spake2021}. In all six cases, the AI-based optimization more precisely constrained atmospheric \ce{H2O} feature amplitudes than previous analyses.}
\label{table:H2O}
\hspace{-2.2cm}
\scalebox{0.88}{%
\begin{tabular}{cccccccccc}
\textbf{Planet} & \textbf{$R_p$} & \textbf{$M_p$} & \textbf{$T_{eq}$ ($K$)} & \textbf{$H_20-J$ ($H$)} & \textbf{$err_{neg}$ ($H$)} & \textbf{$err_{pos}$ ($H$)} & \textbf{$J$ ($\mu m$)} & \textbf{$H_20-J$ ($\mu m$)} & \textbf{Reference}       \
      \\ \hline
GJ-3470 b       & 0.41                        & 0.04                        & 683              & 0.42               & 0.16                    & 0.16                    & 1.240–1.300     & 1.360–1.440       & This work                \\
HAT-P-12 b*     & 0.96                        & 0.21                        & 957              & 0.44               & 0.25                    & 0.25                    & 1.240–1.300     & 1.360–1.440       & This work                \\
HAT-P-17 b*      & 1.05                        & 0.58                        & 794              & 0.07               & 0.69                    & 0.71                    & 1.240–1.300     & 1.360–1.440       & This work                \\
HAT-P-18 b*      & 1.00                        & 0.20                        & 848              & 0.61               & 0.20                    & 0.20                    & 1.240–1.300     & 1.360–1.440       & This work                \\
HAT-P-26 b      & 0.63                        & 0.07                        & 1044             & 1.56               & 0.30                    & 0.30                    & 1.240–1.300     & 1.360–1.440       & This work                \\
HAT-P-41 b      & 2.05                        & 1.19                        & 2138             & -0.09              & 0.47                    & 0.47                    & 1.240–1.300     & 1.360–1.440       & This work                \\
HD-3167 c       & 0.27                        & 0.04                        & 565              & -0.83              & 0.87                    & 0.88                    & 1.240–1.300     & 1.360–1.440       & This work                \\
HD-106315 c     & 0.39                        & 0.05                        & 888              & 2.50               & 0.55                    & 0.55                    & 1.240–1.300     & 1.360–1.440       & This work                \\
HD-149026 b     & 0.74                        & 0.38                        & 1694             & 0.32               & 0.57                    & 0.59                    & 1.240–1.300     & 1.360–1.440       & This work                \\
HIP-41378 b     & 0.22                        & 0.02                        & 912              & -2.65              & 1.59                    & 1.58                    & 1.240–1.300     & 1.360–1.440       & This work                \\
K2-18 b         & 0.21                        & 0.03                        & 282              & 1.05               & 0.67                    & 0.63                    & 1.240–1.300     & 1.360–1.440       & This work                \\
KELT-7 b        & 1.60                        & 1.39                        & 2090             & -0.02              & 0.44                    & 0.45                    & 1.240–1.300     & 1.360–1.440       & This work                \\
Kepler-138 d    & 0.14                        & 0.01                        & 379              & 5.78               & 2.28                    & 2.66                    & 1.240–1.300     & 1.360–1.440       & This work                \\
WASP-29 b*       & 0.77                        & 0.23                        & 962              & 0.10               & 0.42                    & 0.43                    & 1.240–1.300     & 1.360–1.440       & This work                \\
WASP-31 b*      & 1.55                        & 0.48                        & 1574             & 1.13               & 0.39                    & 0.38                    & 1.240–1.300     & 1.360–1.440       & This work                \\
WASP-39 b*      & 1.27                        & 0.28                        & 1121             & 1.22               & 0.16                    & 0.16                    & 1.240–1.300     & 1.360–1.440       & This work                \\
WASP-67 b       & 1.15                        & 0.43                        & 1034             & 0.67               & 0.91                    & 0.92                    & 1.240–1.300     & 1.360–1.440       & This work                \\
WASP-69 b       & 1.11                        & 0.29                        & 988              & 0.38               & 0.13                    & 0.13                    & 1.240–1.300     & 1.360–1.440       & This work                \\
WASP-74 b       & 1.36                        & 0.72                        & 1855             & 0.50               & 0.39                    & 0.39                    & 1.240–1.300     & 1.360–1.440       & This work                \\
WASP-80 b       & 1.00                        & 0.54                        & 827              & 0.40               & 0.27                    & 0.26                    & 1.240–1.300     & 1.360–1.440       & This work                \\
GJ-1214 b       & 0.25                        & 0.03                        & 559              & 0.02               & 0.10                    & 0.10                    & 1.228–1.297     & 1.366–1.435       & \citet{Kreidberg2014b} \\
GJ-436 b        & 0.37                        & 0.07                        & 649              & 0.54               & 0.46                    & 0.46                    & 1.230–1.289     & 1.362–1.438       & \citet{Knutson2014a}   \\
HAT-P-1 b       & 1.32                        & 0.53                        & 1303             & 2.13               & 0.61                    & 0.61                    & 1.223–1.300     & 1.376–1.434       & \citet{Wakeford2013}  \\
HAT-P-3 b       & {\color[HTML]{333333} 0.94} & {\color[HTML]{333333} 0.65} & 1127             & 0.52               & 0.74                    & 0.74                    & 1.220-1.300     & 1.360-1.440       & \citet{Tsiaras2017}    \\
HAT-P-11 b      & 0.39                        & 0.08                        & 870              & 3.47               & 0.72                    & 0.72                    & 1.228–1.303     & 1.360–1.435       & \citet{Fraine2014}     \\
HAT-P-12 b*     & 0.96                        & 0.21                        & 957              & 0.21               & 0.60                    & 0.60                    & 1.226–1.297     & 1.367–1.438       & \citet{Line2013}       \\
HAT-P-17 b*      & 1.05                        & 0.58                        & 780              & 0.27               & 0.78                    & 0.78                    & 1.220-1.300     & 1.360-1.440       & \citet{Tsiaras2017}    \\
HAT-P-18 b*      & 1.00                        & 0.20                        & 843              & 0.51               & 0.28                    & 0.28                    & 1.220-1.300     & 1.360-1.440       & \citet{Hartman2010}    \\
HAT-P-26 b      & 0.63                        & 0.07                        & 980              & 1.92               & 0.31                    & 0.31                    & 1.220-1.300     & 1.360-1.440       & \citet{Wakeford2017_HATP26}    \\
HAT-P-32 b      & {\color[HTML]{333333} 1.98} & {\color[HTML]{333333} 0.68} & 1784             & 1.3                & 0.28                    & 0.28                    & 1.220-1.300     & 1.360-1.440       & \citet{Alam2020}    \\
HAT-P-38 b      & {\color[HTML]{333333} 0.83} & {\color[HTML]{333333} 0.27} & 1080             & 2.03               & 0.66                    & 0.66                    & 1.220-1.300     & 1.360-1.440       & \citet{Sato2012}       \\
HD-97658 b      & 0.19                        & 0.03                        & 733              & 1.85               & 0.91                    & 0.91                    & 1.237–1.292     & 1.366–1.440       & \citet{Knutson2014b}   \\
HD-189733 b     & 1.13                        & 1.13                        & 1199             & 1.86               & 0.36                    & 0.36                    & 1.222–1.297     & 1.372–1.447       & \citet{Mccullough2014} \\
HD-209458 b     & 1.39                        & 0.73                        & 1445             & 1.15               & 0.13                    & 0.13                    & 1.232–1.288     & 1.364–1.439       & \citet{Deming2013}     \\
WASP-12 b       & 1.94                        & 1.47                        & 2581             & 1.48               & 0.32                    & 0.32                    & 1.251–1.320     & 1.389–1.458       & \citet{Kreidberg2015}  \\
WASP-17 b       & 1.87                        & 0.78                        & 1547             & 0.67               & 0.29                    & 0.29                    & 1.240–1.296     & 1.381–1.437       & \citet{Mandell2013}    \\
WASP-19 b       & 1.42                        & 1.15                        & 2064             & 3.03               & 0.64                    & 0.64                    & 1.230–1.286     & 1.371–1.427       & \citet{Mandell2013}    \\
WASP-29 b*       & {\color[HTML]{333333} 0.77} & {\color[HTML]{333333} 0.23} & 963              & 0.12               & 0.49                    & 0.49                    & 1.220-1.300     & 1.360-1.440       & \citet{Tsiaras2017}      \\
WASP-31 b*      & 1.55                        & 0.48                        & 1573             & 0.86               & 0.48                    & 0.48                    & 1.234–1.294     & 1.374–1.434       & \citet{Sing2016}       \\
WASP-39 b*      & 1.27                        & 0.28                        & 1119             & 1.22               & 0.16                    & 0.16                    & 1.220-1.300     & 1.360-1.440       & \citet{Wakeford2018}    \\
WASP-43 b       & 0.93                        & 1.78                        & 1374             & 1.08               & 0.50                    & 0.50                    & 1.228–1.297     & 1.366–1.435       & \citet{Kreidberg2014a} \\
WASP-52 b       & {\color[HTML]{333333} 1.27} & {\color[HTML]{333333} 0.46} & 1300             & 1.33               & 0.28                    & 0.28                    & 1.220-1.300     & 1.360-1.440       & \citet{Bruno2019}    \\
WASP-63 b       & {\color[HTML]{333333} 1.41} & {\color[HTML]{333333} 0.37} & 1508             & 0.39               & 0.30                    & 0.30                    & 1.220-1.300     & 1.360-1.440       & \citet{Tsiaras2017}    \\
WASP-76 b       & {\color[HTML]{333333} 1.83} & {\color[HTML]{333333} 0.92} & 2206             & 1.62               & 0.21                    & 0.21                    & 1.220-1.300     & 1.360-1.440       & \citet{Tsiaras2017}    \\
WASP-96 b       & {\color[HTML]{333333} 1.20} & {\color[HTML]{333333} 0.48} & 1286             & 2.45               & 0.72                    & 0.72                    & 1.226-1.291     & 1.365-1.440       & \citet{Nikolov2022_WASP96b}    \\
WASP-101 b      & {\color[HTML]{333333} 1.43} & {\color[HTML]{333333} 0.51} & 1552             & 0.13               & 0.27                    & 0.27                    & 1.220-1.300     & 1.360-1.440       & \citet{Wakeford2017}    \\
WASP-121 b      & 1.75                        & {\color[HTML]{333333} 1.16} & 2358             & 2.31               & 0.41                    & 0.41                    & 1.220-1.300     & 1.360-1.440       & \citet{Delrez2016}     \\
WASP-127 b      & 1.31                        & {\color[HTML]{333333} 0.16} & 1400             & 0.86               & 0.16                    & 0.17                    & 1.257-1.307     & 1.344-1.492       & \citet{Spake2021}     \\
XO-1 b          & 1.14                        & 0.83                        & 1206             & 1.66               & 0.55                    & 0.55                    & 1.234–1.290     & 1.365–1.422       & \citet{Deming2013}    
\end{tabular}

}

\begin{tablenotes}\footnotesize
\item[*] *Targets repeated in this work and previous works.
\end{tablenotes}

\end{table*}

In addition to standardizing measurements, AI-enabled science offers studies of exoplanet populations on a scale not achievable by humans. This comprehensive approach to population surveys is crucial to comparative exoplanetology studies, with as many samples as possible needed to infer concrete demographic trends. Using the survey results generated via the \texttt{Eureka$!$} HST data optimizer in conjunction with notable previous surveys, this work demonstrates the impact of AI-enabled surveys. These findings, shown in \autoref{fig:plots_Jovian_SubNeptune}, present new, distinctly-separate trends for aerosol-formation between two different classes of planets: Jovians and sub-Neptunes.

It should be noted while the sub-Neptune group of exoplanets shown in this work contains planets with masses both greater and less than Neptune, these planets are referred to as sub-Neptunes for simplicity. 

The modeling study from \citet{Gao2020} of Jovian exoplanet atmospheres identified and modeled the relationship between the planetary equilibrium temperature and water feature amplitude for various surface gravities. This impactful finding established a new understanding for aerosol-formation in Jovian exoplanets. These aerosol models from \citet{Gao2020} for clear and cloudy Jovian atmospheres are included in the upper plot of \autoref{fig:plots_Jovian_SubNeptune} for reference. After more than doubling the number of exoplanets surveyed in \citet{Gao2020}, our findings confirm the accuracy of these Jovian aerosol models. 


Analyzing the survey dataset as a function of mass, a separate pattern for aerosol-formation in sub-Neptunes begins to appear in \autoref{fig:plots_Jovian_SubNeptune}. The lower plot of this figure illustrates this definitively, where Jovian exoplanets are shown above the dotted boundary region ($\sim$ $0.1$ $-$ $0.13$ $M_J$), and sub-Neptunes are shown below. Comparing these \ce{H2O} feature sizes for Jovians and sub-Neptunes on the same plot, we identify trends with planet mass and temperature in the Jovian and sub-Neptune regimes.  Specifically, for Jovians between 700 -- 1500 K, the water band amplitude peaks at cooler temperatures for less massive planets. A similar, colder trend for the sub-Neptune regime of exoplanets is also evident for planets below the dotted boundary region in \autoref{fig:plots_Jovian_SubNeptune}. 


Tying together these observed behavioral similarities spanning from hot Jupiters to sub-Neptunes, our survey presents a prominent demographic trend for successful detection of \ce{H2O} species (water-band amplitude $\geq$ 1). This region is shown in the lower plot of \autoref{fig:plots_Jovian_SubNeptune} as a yellow-shaded region, labeled the “Clear Sky Corridor". The bounds for this region are motivated by both aerosol theory and the empirical data from these planets.  Aerosol theory postulates the hotter, right edge of the corridor
is caused by silicate condensation, while the cooler, left edge is due to haze production \citep{Gao2020}. The suppression due to silicate condensation likely decreases with
temperature as these clouds sink below observable levels, as analogous to the L-T transition
in brown dwarfs \citep{Lueber2022}, while haze begins to dominate at even lower temperatures as the atmospheres contain increasing amounts of methane, a crucial driver of photochemical haze
formation \citep{Moran2020, Gao2021}. While the absence of aerosol in this corridor is apparent, further validation is required to confirm the chemical composition of these species. 

Several previous studies have identified similar demographic regions where exoplanet atmospheres are relatively clear, supporting the concept of our “Clear Sky Corridor". For instance, \cite{Stevenson2016} conducted an archival survey of HST WFC3 observations and found that hot Jupiters with equilibrium temperatures greater than 700 K and logarithmic surface gravities greater than 2.8 ($log(g)$ $>$ 2.8) display stronger water absorption features, suggesting reduced cloud opacity at higher temperatures. Likewise, \cite{Fu2017} explored the influence of equilibrium temperature and metallicity on cloud formation, demonstrating that planets within certain temperature ranges are less likely to develop high-altitude clouds, thereby enhancing the detectability of molecular features. \cite{Crossfield2017} reviewed transmission spectroscopy results and noted a trend where planets with specific temperature and gravity combinations exhibit more pronounced spectral features, consistent with fewer obscuring aerosols. More recently, \cite{Dymont2022} and \cite{Brande2024} expanded upon these findings using larger datasets and refined models, further delineating the regions of parameter space associated with clear atmospheres. Our identification of the Clear Sky Corridor aligns with these studies but distinguishes itself by applying AI-driven analysis to a broader dataset, enhancing the statistical robustness of this demographic trend. By contextualizing our results within this continuum of research, we underscore the pivotal role of temperature-dependent aerosol processes in determining atmospheric clarity and the potential for successful atmospheric characterization.

The atmosphere of exoplanet \textit{K2-18 b} has been surveyed as part of this work, with results detailed in \autoref{table:H2O}. However, while our HST WFC3 analysis supports historical detections of \ce{H2O} within the $1.36-1.44$ $\mu m$ near-infrared band (\citealt{Benneke2017, Benneke2019}), follow-up JWST observations determined this molecular absorption was due to the presence of methane (\citealt{Madhusudhan2023}). For this reason, \textit{K2-18 b} has been omitted from the water-band amplitude measurements shown in \autoref{fig:plots_Jovian_SubNeptune}. It is worth noting that \textit{K2-18 b} is significantly cooler than planets shown in \autoref{fig:plots_Jovian_SubNeptune}, and would fall well outside of this “Clear Sky Corridor". While this demographic corridor for atmospheric observation of exoplanets appears real, follow-up JWST observations are required to resolve this ambiguity and validate the chemical composition of these detected atmospheric species.

Through this large-scale, data-driven exercise in comparative exoplanetology, we have uncovered significant trends in aerosol formation across different exoplanet types. Our findings not only strongly corroborate the modeled relationship between water-band amplitude and planet temperature for hot Jupiters identified by \citet{Gao2020}, but also excitingly reveal a similar, but novel trend in \ce{H2O} feature amplitude for sub-Neptunes. These insights are pivotal in enhancing our understanding of exoplanet atmospheres, specifically as it relates to metallicity and aerosol formation. By integrating these new discoveries into our models of planetary aerosol formation, we are poised to advance the study of exoplanet atmospheres. The foundational knowledge provided by HST sets the stage for JWST to explore similar trends across broader wavelengths and for additional chemical species, promising a bright future for the study of exoplanet atmospheres.


A more comprehensive survey will further clarify this new understanding of aerosol formation across exoplanet-types, but this novel finding is exciting and long-awaited. These patterns of atmospheric composition for different classes of exoplanets were notably suggested by \citet{Fortney2008}, and \citet{Seager2010}. Now, with cutting-edge software maximizing the capabilities of state-of-the-art facilities, we are beginning to advance into these next stages of exoplanet discovery.

\section{Summary}
\label{sec:Summary}

As teams lead observations of exoplanet atmospheres, the standard process is to analyze the data, perform retrievals, and then publish the results. Often, the bottlenecks in this flow are the nuanced methods associated with reducing the data into light curves and fitting the instrument systematics. 

AI-based optimization of these procedures is a novel development in the field, as no other software in the scientific community offers such automation for processing transit observations of exoplanet atmospheres. Automating this optimization process significantly reduces hours otherwise spent processing HST observations.

Using the \texttt{Eureka$!$} AI-based HST optimizer presented in this work, we provide one of the most comprehensive surveys of exoplanet atmospheres to-date, identifying significant trends in aerosol formation across different planet regimes (shown in \autoref{fig:plots_Jovian_SubNeptune}). Using the \ce{H2O}-J metric to measure spectral features, we have successfully constrained water feature sizes for the atmospheres of exoplanets surveyed in this work, and identified patterns for aerosol formation across different planet regimes: Jovians and sub-Neptunes. Building on this foundation, we will extend AI-based optimization to JWST's NIRCam, NIRSpec, NIRISS, and MIRI-LRS modes with \texttt{Eureka$!$} in future work, allowing us to exploit higher quality data over a greater wavelength range to assess additional exoplanet population trends.


The rapid and reliable science enabled by AI-based surveys will increase the rate of scientific discovery in the field of exoplanets. The results of this initial AI-based survey, spanning the Jovian and sub-Neptune planet regimes, already presents significant findings. Expanding investigation of the “Clear Sky Corridor" presented in this work, JWST observations will provide necessary answers to better bound this newly-found demographic region. Importantly, follow-up observations with JWST will reveal how persistent these patterns for aerosol formation are across a broader range of planets and wavelengths. 

AI-based telescope data processing will inevitably increase the rate of discovery in the field of exoplanet atmospheres. With more rapid and reliable exoplanet atmospheric characterization, this exciting advancement will simplify, accelerate, and advance how we utilize our most valuable observatories. This technology development and the unbiased, trusted look it provides of exoplanet atmospheres is an obvious need for observers, the exoplanet community as a whole, and across all of NASA.

\section*{Acknowledgements}
The authors would like to thank Laura Mayorga, Arika Egan, Erin May, Jacob Lustig-Yaeger, Hannah Wakeford, Sarah Moran, Guangwei Fu and Peter Gao for their contributions to this work. 

This research was supported by NASA through grants under the HST-AR-16634 program from STScI. 

This research is based on observations made with the NASA/ESA Hubble Space Telescope obtained from the Space Telescope Science Institute (STScI), which is operated by the Association of Universities for Research in Astronomy, Inc., under NASA contract NAS 5-26555. 

This research has made use of NASA's Astrophysics Data System Bibliographic Services and the \cite{exoplanetArchive_PS}, which is operated by the California Institute of Technology, under contract with NASA under the Exoplanet Exploration Program.

Some of the data presented in this article was obtained from the Mikulski Archive for Space Telescopes (MAST) at the STScI. The specific observations analyzed can be accessed via \dataset[DOI]{https://doi.org/10.17909/thfs-g992}.



\section*{Appendix}
\label{sec:appendix}

\setcounter{table}{0}
\renewcommand{\thetable}{A\arabic{table}}

This appendix contains a detailed description of the data reduction and lightcurve generation parameters of the \texttt{Eureka$!$} program optimized as part of this work. Note that the variables listed here are only the parameters whose values are optimized as part of the AI-based algorithm demonstrated in this publication. Detailed descriptions of all \texttt{Eureka$!$} ECF parameters and further details on how to operate the software are available on the \texttt{Eureka$!$} \texttt{ReadTheDocs} page\footnote[1]{https://eurekadocs.readthedocs.io/en/latest/index.html}.

\autoref{table:optimized_parameters} lists the optimized parameters from the \texttt{Eureka$!$} electronic control files. Additionally, \autoref{table:target_info_sidebyside} provides detailed information about the 20 exoplanets analyzed in this work, including their observation visit numbers, program IDs, years of observation, and principal investigators. Finally, \autoref{table:app_fit_params} presents the MCMC-fitted orbital parameters used to model the white light curves of these transiting exoplanets.


\begin{table*}[] 
\noindent
\resizebox{\textwidth}{!}{%
\begin{tabular}{ll}
\multicolumn{2}{l}{\textbf{Stage 3: Data Reduction}}      
\\
\hline
\hline
\textbf{Parameter}      & \textbf{Description}                                                                                                         \\
\hline

\textit{xwindow}        & X-axis window dimensions, in pixels, for location of spectra on the detector                                                 \\
\textit{ywindow}        & Y-axis window dimensions, in pixels, for location of spectra on the detector                                                 \\
\textit{diffthresh}     & Sigma threshold for bad pixel identification in the differential non-destructive reads                                       \\
\textit{bg\_hw}         & Half-width of exclusion region for background subtraction, in pixels                                                         \\
\textit{bg\_thresh}     & Double-iteration X-sigma threshold for outlier rejection along time axis                                                     \\
\textit{bg\_deg}        & Polynomial order for column-by-column background subtraction, -1 for median of entire frame                                \\
\textit{p3thresh}       & X-sigma threshold for outlier rejection during background subtraction                                                        \\
\textit{spec\_hw}       & Half-width of the aperture region used for spectral extraction, in pixels                                                    \\
\textit{window\_len}    & Smoothing window length for the trace location, in pixels                                                                    \\
\textit{median\_thresh} & Sigma threshold when flagging outliers in median frame                                                                       \\
\textit{p5thresh}       & X-sigma threshold for outlier rejection while constructing spatial profile of the extracted spectra                          \\
\textit{p7thresh}       & X-sigma threshold for outlier rejection during optimal spectral extraction      

\vspace{0.5cm}

\\

\multicolumn{2}{l}{\textbf{Stage 4: Lightcurve Generation}}       \\

\hline
\hline

\textbf{Parameter}      & \textbf{Description}                                                                                                         \\
\hline

\textit{drift\_range}   & Trim spectra by +/-X pixels to compute valid region of cross correlation for 1D spectral drift correction                    \\
\textit{drift\_hw}      & Half-width in pixels used when fitting Gaussian for 1D spectral drift correction                                             \\
\textit{highpassWidth}  & The integer width of the highpass filter when subtracting the continuum                                                      \\
\textit{sigma}          & The number of sigmas a point must be from the rolling median to be considered an outlier and clipped                         \\
\textit{box\_width}     & The width of the box-car filter (used to calculated the rolling median for sigma clipping) in units of number of data points
\end{tabular}
} 
\caption{Parameters from the Eureka! electronic control files optimized using the AI-based algorithm presented in this work.}
\label{table:optimized_parameters}
\end{table*}

\begin{table*}[!]
\resizebox{0.98\textwidth}{!}{%
\hspace{-2.8cm}
\begin{tabular}{ccccc|ccccc}
\textbf{Program ID} & \textbf{Visit} & \textbf{Target} & \textbf{Year} & \textbf{PI} & \textbf{Program ID} & \textbf{Visit} & \textbf{Target} & \textbf{Year} & \textbf{PI} \\ \hline
12473               & 25             & WASP-31 b       & 2012          & Sing        & 14260               & 18             & HAT-P-18 b      & 2016          & Deming      \\
12956               & 2              & HAT-P-17 b      & 2013          & Huitson     & 14260               & 19             & HAT-P-18 b      & 2017          & Deming      \\
13665               & 1              & Kepler-138 d    & 2014          & Benneke     & 14260               & 20             & HAT-P-26 b      & 2016          & Deming      \\
13665               & 2              & Kepler-138 d    & 2015          & Benneke     & 14260               & 21             & HAT-P-26 b      & 2016          & Deming      \\
13665               & 3              & Kepler-138 d    & 2015          & Benneke     & 14260               & 9              & HD-149026 b     & 2016          & Deming      \\
13665               & 24             & GJ-3470 b       & 2015          & Benneke     & 14260               & 15             & WASP-29 b       & 2016          & Deming      \\
13665               & 25             & GJ-3470 b       & 2015          & Benneke     & 14260               & 10             & WASP-39 b       & 2016          & Deming      \\
13665               & 26             & GJ-3470 b       & 2015          & Benneke     & 14260               & 11             & WASP-39 b       & 2017          & Deming      \\
13665               & 29             & K2-18 b         & 2015          & Benneke     & 14260               & 7              & WASP-67 b       & 2016          & Deming      \\
13665               & 30             & K2-18 b         & 2016          & Benneke     & 14260               & 14             & WASP-69 b       & 2016          & Deming      \\
13665               & 35             & K2-18 b         & 2016          & Benneke     & 14260               & 5              & WASP-80 b       & 2016          & Deming      \\
14682               & 1              & K2-18 b         & 2017          & Benneke     & 14767               & 87             & HAT-P-41 b      & 2016          & Sing        \\
14682               & 2              & K2-18 b         & 2017          & Benneke     & 14767               & 91             & KELT-7 b        & 2017          & Sing        \\
14682               & 3              & K2-18 b         & 2016          & Benneke     & 14767               & 79             & WASP-74 b       & 2016          & Sing        \\
14682               & 4              & K2-18 b         & 2017          & Benneke     & 15333               & 11             & HD-106315c      & 2018          & Crossfield  \\
14682               & 5              & K2-18 b         & 2017          & Benneke     & 15333               & 13             & HD-106315c      & 2018          & Crossfield  \\
14260               & 16             & HAT-P-12 b      & 2015          & Deming      & 15333               & 14             & HD-106315c      & 2019          & Crossfield  \\
14260               & 17             & HAT-P-12 b      & 2016          & Deming      & 15333               & 48             & HD-106315c      & 2019          & Crossfield  \\
14099               & 4              & HAT-P-18 b      & 2015          & Evans       & 15333               & 1              & HD-3167c        & 2018          & Crossfield  \\
                   &                &                 &               &             & 15333               & 3              & HD-3167c        & 2018          & Crossfield  \\
                   &                &                 &               &             & 15333               & 5              & HD-3167c        & 2019          & Crossfield  \\
                   &                &                 &               &             & 15333               & 47             & HD-3167c        & 2019          & Crossfield  \\
                   &                &                 &               &             & 15333               & 21             & HIP-41378 b     & 2018          & Crossfield  \\
                   &                &                 &               &             & 15333               & 22             & HIP-41378 b     & 2018          & Crossfield  \\
                   &                &                 &               &             & 15333               & 56             & HIP-41378 b     & 2020          & Crossfield  
\end{tabular}%
}
\caption{List of 20 exoplanets analyzed in this work, along with the corresponding observation visit number(s), program ID, year of observation, and principal investigator for each target.}
\label{table:target_info_sidebyside}
\end{table*}

\begin{table*}[ht!]
\caption{MCMC-fitted orbital parameters used to model the white light curves of the 20 transiting exoplanets analyzed in this work.}
\scalebox{1.18}{
\hspace{-2cm}
\begin{tabular}{clllllllll}
\textbf{Planet} & \multicolumn{1}{c}{\textbf{$R_p/R_s$}} & \multicolumn{1}{c}{\textbf{P (days)}} & \multicolumn{1}{c}{\textbf{$t_0$ (MJD)}} & \multicolumn{1}{c}{\textbf{i ($^{\circ}$)}} & \multicolumn{1}{c}{\textbf{a/Rs}} & \multicolumn{1}{c}{\textbf{e}} & \multicolumn{1}{c}{\textbf{w ($^{\circ}$)}} & \multicolumn{1}{c}{\textbf{$u_1$}} & \multicolumn{1}{c}{\textbf{$u_2$}} \\ \hline
GJ-3470 b            & 0.1451                             & 4.055                                 & 56400.9037                            & 87.74                                & 11.3695                           & 0                              & 90.0                                 & 0.17                            & 0.25                            \\
HAT-P-12 b           & 0.1389                             & 3.213                                 & 56716.0305                            & 87.61                                & 10.7052                           & 0                              & 354.3                                & 0.24                            & 0.26                            \\
HAT-P-17 b           & 0.1217                             & 10.339                                & 56568.5586                            & 89.37                                & 22.9118                           & 0.52                           & 203.1                                & 0.21                            & 0.26                            \\
HAT-P-18 b           & 0.1364                             & 5.508                                 & 59743.3534                            & 88.62                                & 17.4922                           & 0.08                           & 12.0                                 & 0.21                            & 0.26                            \\
HAT-P-26 b           & 0.0738                             & 4.235                                 & 56892.0904                            & 85.91                                & 10.4460                           & 0.12                           & 46.0                                 & 0.21                            & 0.26                            \\
HAT-P-41 b           & 0.0990                             & 2.694                                 & 58070.7439                            & 85.87                                & 5.1360                            & 0                              & 90.0                                 & 0.12                            & 0.20                            \\
HD-106315c           & 0.0287                             & 21.057                                & 57610.6209                            & 89.40                                & 28.2977                           & 0.22                           & 96.0                                 & 0.12                            & 0.20                            \\
HD-149026 b          & 0.0518                             & 2.876                                 & 57217.1415                            & 84.50                                & 5.7816                            & 0                              & 109.0                                & 0.15                            & 0.23                            \\
HD-3167c             & 0.0328                             & 29.846                                & 58439.1081                            & 89.47                                & 43.7012                           & 0.15                           & 178.0                                & 0.17                            & 0.25                            \\
HIP-41378 b          & 0.0172                             & 15.572                                & 57151.7998                            & 88.80                                & 19.0000                           & 0                              & 90.0                                 & 0.15                            & 0.23                            \\
K2-18 b              & 0.0524         & 32.940            & 57725.0483        & 89.58            & 78.1429       & 0.2        & -5.7             & 0.29        & 0.22        \\
KELT-7 b             & 0.0881                             & 2.735                                 & 59934.5397                            & 83.70                                & 5.2451                            & 0                              & 90.0                                 & 0.11                            & 0.20                            \\
Kepler-138 d         & 0.0246                             & 23.093                                & 57013.2478                            & 89.04                                & 52.7900                           & 0.02                           & 34.0                                 & 0.29                            & 0.22                            \\
WASP-29 b            & 0.0971                             & 3.923                                 & 58355.9151                            & 89.65                                & 11.5473                           & 0                              & 90.0                                 & 0.21                            & 0.26                            \\
WASP-31 b            & 0.1238                             & 3.406                                 & 55873.3666                            & 84.36                                & 8.1266                            & 0.30                           & 14.9                                 & 0.12                            & 0.22                            \\
WASP-39 b            & 0.1451                             & 4.055                                 & 56400.9037                            & 87.74                                & 11.3695                           & 0                              & 90.0                                 & 0.17                            & 0.25                            \\
WASP-67 b            & 0.1408                             & 4.614                                 & 56617.5533                            & 85.95                                & 12.6551                           & 0                              & 335.4                                & 0.21                            & 0.26                            \\
WASP-69 b            & 0.1253                             & 3.868                                 & 59798.2758                            & 87.67                                & 13.4334                           & 0                              & 90.0                                 & 0.24                            & 0.26                            \\
WASP-74 b            & 0.0954                             & 2.138                                 & 57205.4378                            & 79.89                                & 4.9150                            & 0                              & 90.0                                 & 0.15                            & 0.23                            \\
WASP-80 b            & 0.1687                             & 3.068                                 & 56486.925                             & 89.90                                & 11.6753                           & 0                              & 94.0                                 & 0.29                            & 0.22                            \\
\end{tabular}
\label{table:app_fit_params}
}
\end{table*}


\clearpage

\bibliography{HST_SHEL_WFC3_Survey}{}
\bibliographystyle{aasjournal}

\end{document}